\documentclass{aa}  
\usepackage[]{aalongtable}
\usepackage{scalefnt}
\usepackage{pdflscape}
\usepackage{graphicx}
\usepackage{txfonts}
\def\kms   {km~s$^{-1}$}
\def\loggf {log~$gf$}
\def\Teff  {$T_\mathrm{eff}$}
\def\logg  {log~$g$}
\begin{document}

\title{Looking for imprints of the first stellar generations in metal-poor bulge field stars
\thanks{Observations collected at the European Southern Observatory,
Paranal, Chile (ESO), under programmes 089.B-0208(A).}}


\author{
C. Siqueira-Mello\inst{1}
\and
C Chiappini\inst{2}
\and
B. Barbuy\inst{1}
\and
K. Freeman\inst{3}
\and
M. Ness\inst{4}
\and
E. Depagne\inst{5}
\and
E. Cantelli\inst{1}
\and
M. Pignatari\inst{6,7,8}
\and
R. Hirschi\inst{9,10}
\and
U. Frischknecht\inst{9,11}
\and
G. Meynet\inst{12}
\and
A. Maeder\inst{12}
}
\offprints{C. Siqueira Mello Jr. (cesar.mello@usp.br).}

\institute{
Universidade de S\~ao Paulo, IAG, Rua do Mat\~ao 1226,
Cidade Universit\'aria, S\~ao Paulo 05508-900, Brazil;
e-mail: cesar.mello@usp.br
\and
Leibniz-Institut f\"ur Astrophysik Potsdam (AIP), 
An der Sternwarte 16, 14482, Potsdam, Germany;
e-mail: cristina.chiappini@aip.de
\and
Mount Stromlo Observatory, Research School of Astronomy \& Astrophysics, Australian 
National University College of Physical and Mathematical Sciences, 
 Cotter Road. Weston Creek, ACT 2611, Australia 
\and
Max Planck Institute for Astronomy, K\"onigstuhl 17,  69117 Heidelberg, Germany
\and
South African Astronomical Observatory (SAAO), P.O. Box 9. Observatory 7935, South Africa
\and 
E.A. Milne Centre for Astrophysics, Dept of Physics \& Mathematics, University of Hull, HU6 7RX, United Kingdom
\and 
Konkoly Observatory, Research Centre for Astronomy and Earth Sciences, Hungarian Academy of Sciences, 
Konkoly Thege Miklos ut 15-17, H-1121 Budapest, Hungary
\and
NuGrid collaboration, \url{http://www.nugridstars.org}
\and
Astrophysics group, Lennard-Jones Laboratories, Keele University, ST5 5BG, Staffordshire, UK
\and
Kavli Institute for the Physics and Mathematics of the Universe (WPI), University of Tokyo, 5-1-5 Kashiwanoha, 
Kashiwa, 277-8583, Japan
\and
Dept. of Physics, University of Basel, Klingelbergstr. 82, 4056, Basel, Switzerland
\and
Observatoire de Gen\`eve, Chemin des Maillettes 51, Sauverny, CH-1290 Versoix, Switzerland 
}
\date{Received; accepted}

\titlerunning{}
\abstract
{Efforts to look for signatures of the first stars have concentrated on metal-poor halo objects. 
However, the low end of the bulge metallicity distribution has been shown to host some of the 
oldest objects in the Milky Way and hence this Galactic component potentially offers interesting 
targets to look at imprints of the first stellar generations. As a pilot project, 
we selected bulge field stars already identified in the ARGOS survey as having 
[Fe/H]~$\approx-1$ and oversolar [$\alpha$/Fe] ratios, and we used FLAMES-UVES to obtain detailed 
abundances of key elements that are believed to reveal imprints of the first stellar generations.}
{The main purpose of this study is to analyse selected ARGOS stars using new high-resolution 
(R$\sim$45,000) and high-signal-to-noise (S/N$>$100) spectra. We aim to derive their stellar 
parameters and elemental ratios, in particular the abundances of C, N, the $\alpha$-elements 
O, Mg, Si, Ca, and Ti, the odd-Z elements Na and Al, the neutron-capture s-process dominated 
elements Y, Zr, La, and Ba, and the r-element Eu.}
{High-resolution spectra of five field giant stars were obtained at the 8m VLT UT2-Kueyen 
telescope with the UVES spectrograph in FLAMES-UVES configuration. Spectroscopic parameters 
were derived based on the excitation and ionization equilibrium of \ion{Fe}{I} and \ion{Fe}{II}. 
The abundance analysis was performed with a MARCS LTE spherical model atmosphere grid and the 
Turbospectrum spectrum synthesis code.}
{We confirm that the analysed stars are moderately metal-poor ($-1.04\leq$~[Fe/H]~$\leq-0.43$), 
non-carbon-enhanced (non-CEMP) with [C/Fe]~$\leq+0.2$, and $\alpha$-enhanced. We find that our three 
most metal-poor stars are nitrogen enhanced. The $\alpha$-enhancement suggests that these stars were 
formed from a gas enriched by core-collapse supernovae, and that the values are in agreement with 
results in the literature for bulge stars in the same metallicity range. No abundance anomalies 
(Na~$-$~O, Al~$-$~O, Al~$-$~Mg anti-correlations) were detected in our sample. 
The heavy elements Y, Zr, Ba, La, and Eu also exhibit oversolar abundances. 
Three out of the five stars analysed here show slightly enhanced [Y/Ba] ratios similar to 
those found in two other metal-poor bulge GCs (NGC~6522 and M~62). This sample shows enhancement in the 
first-to-second peak abundance ratios of heavy elements, as well as dominantly s-process element excesses. 
This can be explained by different nucleosynthesis scenarios.}
{
}

\maketitle

%

\section{Introduction}
 
Efforts to find the chemical imprints in the oldest stars of the Milky Way left by the first 
stellar generations (hereafter, first stars) have focused on very metal-poor halo stars 
with [Fe/H]~$\sim-3$. Some cosmological simulations have suggested that at least half 
of the first stars should have formed in the Galactic bulge (e.g. Tumlinson 2010). 
Consequently, this Galactic component is a potential source of interesting targets to be explored. 
These simulations suggest that the oldest stars, which have formed at the highest density peaks 
(bulge), have enriched the surrounding interstellar medium (ISM) on very short timescales. 
Chemical evolution models also suggest that the old bulge formed on short timescales 
(e.g. Grieco et al. 2012). 

Barbuy et al. (2009, 2014) searched for evidence of the signatures of formation of the first 
stellar generations in the old bulge globular cluster NGC 6522, potentially the oldest Milky Way 
globular cluster. The results were discussed in the framework of the early fast-rotating massive 
stars, coined Spinstars, or mass transfer from AGB stars (Chiappini et al. 2011; Ness et al. 2014). 

The central parts of the Galaxy, where the oldest stars most probably preferentially reside, 
have not been targeted extensively, partly due to the very small fraction of metal-poor stars 
in the predominantly metal-rich bulge region. The situation started to change with the Bulge Radial 
Velocity Assay (BRAVA; Kunder et al. 2012) and the Abundances and Radial velocity Galactic Origins 
Survey (ARGOS; Freeman et al. 2013), and the new data being obtained by the Apache Point Observatory 
Galactic Evolution Experiment (APOGEE; Majewski et al. 2015) in the near infrared. In particular, 
the Galactic bulge ARGOS Survey, an AAOmega/AAT spectroscopic survey that measured radial 
velocities, metallicities and [$\alpha$/Fe] ratio of about 28, 000 stars (Freeman et al. 2013), 
opened the opportunity to explore the bulge field metal-poor stars. The rather large number of 
targets observed by ARGOS provided the opportunity of identifying bulge stars with metallicities 
[Fe/H]~$\approx-1$, and estimating oversolar [$\alpha$/Fe] ratios. 

Using ARGOS targets, we began a pilot project aimed at obtaining detailed chemical abundances of 
metal-poor bulge field stars to look for possible chemical imprints of the first stars. 
One of these imprints could be an overabundance of dominantly s-process elements (e.g. Meynet et al. 2006; 
Pignatari et al. 2008; Chiappini et al. 2011; Frischknecht et al. 2012, 2015; 
Barbuy et al. 2014; Cecutti et al. 2013; Cescutti \& Chiappini 2014). 
Indeed, enhancements of Sr, Y, and Zr relative to Ba and La, and an excess of Ba or La relative to Eu, 
in very old stars of the Milky Way can be attributed to the s-process activation in early generations of 
fast rotating massive stars, which pollute the primordial material prior to the formation of the oldest 
bulge (halo) field stars. An alternative possibility is an s-process contribution from massive AGB stars 
bound in a binary system (e.g. Bisterzo et al. 2010). Otherwise the idea that has been more widely accepted 
is that these elements were produced by the r-process in early times (Truran 1981). An extra process is also 
claimed to produce the enhancement of the lightest heavy elements relative to the heaviest elements, in 
literature called the lighter element primary process (LEPP; Travaglio et al. 2004; Bisterzo et al. 2014) 
or weak r-process (Wanajo \& Ishimaru 2006). The astrophysical scenarios with neutrino-driven winds are 
considered the most promising sites (Wanajo 2013; Arcones \& Thielemann 2013; Fujibayashi et al. 2015; 
Niu et al. 2015).

The aim of this work is to obtain detailed chemical constraints from field bulge stars. Here, we analyse five 
of these stars at high spectral resolution, using UVES spectra. We derive element abundances of C, N, 
the $\alpha$-elements O, Mg, Si, Ca, and Ti, the odd-Z elements Na and Al, the dominantly s-elements Y, Zr, Ba, 
and La, and the r-element Eu. The observations are described in Sect. 2. Photometric effective temperatures are 
derived in Sect. 3. Spectroscopic parameters are derived in Section 4 and abundance ratios are computed in Sect. 5. 
A discussion is presented in Sect. 6 and conclusions are drawn in Sect. 7.

\begin{table}
\caption{Log of the spectroscopic observations: date, time, exposure time, 
seeing, and air mass 
at the beginning and at the end of the observation.}             
\label{log}      
\scalefont{0.87}
\centering                          
\begin{tabular}{cccccc}        
\hline\hline                 
\noalign{\smallskip}
\hbox{Run} & \hbox{Date} & \hbox{Time} &\hbox{Exp.} &\hbox{Seeing} &\hbox{Airmass} \\
\noalign{\smallskip}
\hline
\noalign{\smallskip}
\hbox{} & \hbox{} & \hbox{}  & \hbox{(s)} & \hbox{(``)} & \hbox{} \\
\noalign{\smallskip}
\hline
\noalign{\smallskip}
\hbox{1}  & 2012-07-12 & 04:07:43.1 & 2775 & 0.8$-$0.8 & 1.0$-$1.1\\
\hbox{2}  & 2012-08-02 & 02:59:47.1 & 2775 & 0.3$-$0.7 & 1.0$-$1.1\\
\hbox{3}  & 2012-07-21 & 03:59:12.6 & 2775 & 1.0$-$1.3 & 1.0$-$1.1\\
\hbox{4}  & 2012-07-23 & 02:46:14.3 & 2775 & 0.7$-$1.0 & 1.0$-$1.0\\
\hbox{5}  & 2012-08-02 & 03:57:27.9 & 2775 & 1.0$-$1.5 & 1.1$-$1.2\\
\hbox{6}  & 2012-08-03 & 23:31:53.7 & 2775 & 0.8$-$0.8 & 1.2$-$1.1\\
\hbox{7}  & 2012-08-22 & 01:22:09.9 & 2775 & 0.7$-$0.8 & 1.0$-$1.1\\
\hbox{8}  & 2012-08-21 & 23:56:35.2 & 2308 & 0.9$-$0.9 & 1.0$-$1.0\\
\hbox{9}  & 2012-08-22 & 02:24:20.8 & 2775 & 0.7$-$0.7 & 1.1$-$1.2\\
\hbox{10} & 2012-08-23 & 01:48:05.3 & 2775 & 1.0$-$1.1 & 1.0$-$1.1\\
\noalign{\smallskip}
\hline                                   
\end{tabular}
\end{table}

\begin{table}[ht!]
\caption{Geocentric radial velocity in each of the ten exposure runs with corresponding 
heliocentric radial velocities and mean heliocentric radial velocity, in \kms.}             
\label{vradial}      
\centering                          
\begin{tabular}{crrrr}        
\hline\hline                 
\noalign{\smallskip}
\hbox{run} & \hbox{RV$\rm_{G}$} & \hbox{RV$\rm_{B}$} & \hbox{RV$\rm_{G}$} & \hbox{RV$\rm_{B}$} \\
\noalign{\smallskip}
\hline
\noalign{\smallskip}
             &\multicolumn{2}{c}{221}&\multicolumn{2}{c}{224}\\
\noalign{\smallskip}
\hline
\noalign{\smallskip}             
\hbox{run1}   & $-$96.9$\pm$2.1 & $-$106.7$\pm$2.1 & $-$105.9$\pm$2.1 & $-$115.6$\pm$2.1 \\
\hbox{run2}   & $-$88.3$\pm$1.6 & $-$107.1$\pm$1.6 &  $-$96.7$\pm$2.0 & $-$115.4$\pm$2.0 \\
\hbox{run3}   & $-$92.4$\pm$1.8 & $-$106.3$\pm$1.8 & $-$101.1$\pm$2.0 & $-$115.0$\pm$2.0 \\
\hbox{run4}   & $-$66.9$\pm$2.1 & $-$81.5$\pm$2.1  & $-$101.6$\pm$3.8 & $-$116.2$\pm$3.8 \\
\hbox{run5}   & $-$87.5$\pm$2.4 & $-$106.3$\pm$2.4 &  $-$96.3$\pm$2.3 & $-$115.1$\pm$2.3 \\
\hbox{run6}   & $-$87.8$\pm$1.4 & $-$106.9$\pm$1.4 &  $-$96.2$\pm$1.8 & $-$115.3$\pm$1.8 \\
\hbox{run7}   & $-$81.9$\pm$2.0 & $-$107.1$\pm$2.0 &  $-$90.4$\pm$2.1 & $-$115.5$\pm$2.1 \\
\hbox{run8}   & $-$81.7$\pm$3.4 & $-$106.7$\pm$3.4 &  $-$89.3$\pm$3.0 & $-$114.3$\pm$3.0 \\
\hbox{run9}   & $-$80.8$\pm$1.6 & $-$106.1$\pm$1.6 &  $-$89.3$\pm$1.6 & $-$114.5$\pm$1.6 \\
\hbox{run10}  & $-$81.9$\pm$1.4 & $-$107.4$\pm$1.4 &  $-$90.2$\pm$1.5 & $-$115.7$\pm$1.5 \\
\noalign{\smallskip}
\hline
\noalign{\smallskip}
\hbox{Mean}  &  & $-$106.7$\pm$2.1 &  & $-$115.3$\pm$2.4 \\
\noalign{\smallskip}
\hline
\noalign{\smallskip}
             &\multicolumn{2}{c}{230}&\multicolumn{2}{c}{235}\\
\noalign{\smallskip}
\hline
\noalign{\smallskip}             
\hbox{run1}   & $-$71.1$\pm$2.1 & $-$80.9$\pm$2.1 & 145.4$\pm$1.4 & 135.6$\pm$1.4 \\
\hbox{run2}   & $-$62.2$\pm$2.1 & $-$80.9$\pm$2.1 & 154.3$\pm$1.6 & 135.6$\pm$1.6 \\
\hbox{run3}   & $-$66.2$\pm$1.9 & $-$80.1$\pm$1.9 & 150.6$\pm$1.6 & 136.7$\pm$1.6 \\
\hbox{run4}   & -------------   & -------------   & 149.7$\pm$1.5 & 135.1$\pm$1.5 \\
\hbox{run5}   & -------------   & -------------   & 155.5$\pm$1.6 & 136.6$\pm$1.6 \\
\hbox{run6}   & $-$61.5$\pm$2.5 & $-$80.6$\pm$2.5 & 155.0$\pm$1.5 & 135.9$\pm$1.5 \\
\hbox{run7}   & $-$55.9$\pm$1.9 & $-$81.0$\pm$1.9 & 160.9$\pm$1.6 & 135.7$\pm$1.6 \\
\hbox{run8}   & $-$54.7$\pm$4.4 & $-$79.7$\pm$4.4 & 162.0$\pm$1.6 & 137.0$\pm$1.6 \\
\hbox{run9}   & $-$54.7$\pm$1.9 & $-$80.0$\pm$1.9 & 162.2$\pm$1.6 & 136.9$\pm$1.6 \\
\hbox{run10}  & $-$55.5$\pm$1.8 & $-$80.9$\pm$1.8 & 161.0$\pm$1.4 & 135.6$\pm$1.4 \\
\noalign{\smallskip}
\hline
\noalign{\smallskip}
\hbox{Mean}  &  & $-$80.5$\pm$2.3 &  & 136.1$\pm$1.6 \\
\noalign{\smallskip}
\hline
\noalign{\smallskip}
             & \multicolumn{2}{c}{238} &  & \\
\hbox{run1}   & $-$137.1$\pm$1.5 & $-$146.9$\pm$1.5 &  & \\
\hbox{run2}   & $-$128.0$\pm$1.6 & $-$146.7$\pm$1.6 &  & \\
\hbox{run3}   & $-$132.3$\pm$1.5 & $-$146.3$\pm$1.5 &  & \\
\hbox{run4}   & $-$132.6$\pm$1.8 & $-$147.3$\pm$1.8 &  & \\
\hbox{run5}   & $-$127.4$\pm$1.6 & $-$146.3$\pm$1.6 &  & \\
\hbox{run6}   & $-$127.5$\pm$1.7 & $-$146.6$\pm$1.7 &  & \\
\hbox{run7}   & $-$121.5$\pm$1.5 & $-$146.7$\pm$1.5 &  & \\
\hbox{run8}   & $-$121.0$\pm$1.9 & $-$146.0$\pm$1.9 &  & \\
\hbox{run9}   & $-$120.8$\pm$1.4 & $-$146.1$\pm$1.4 &  & \\
\hbox{run10}  & $-$121.5$\pm$1.4 & $-$147.0$\pm$1.4 &  & \\
\noalign{\smallskip}
\hline
\noalign{\smallskip}
\hbox{Mean}  &  & $-$146.6$\pm$1.7 &  & \\
\noalign{\smallskip}
\hline                                   
\end{tabular}
\end{table}

\begin{table*}[ht!]
\caption[1]{Identifications, coordinates, magnitudes, and reddening. 
$JHK_{s}$ from both 2MASS and VVV surveys are given. }
\small
\scalefont{0.85}
\begin{flushleft}
\tabcolsep 0.15cm
\begin{tabular}{ccccccrrrrrrrrrr}
\noalign{\smallskip}
\hline
\noalign{\smallskip}
\hline
\noalign{\smallskip}
{\rm star} & 2MASS ID & \hbox{$\alpha$(J2000)} & \hbox{$\delta$(J2000)} & \hbox{l($^{\circ}$)} & \hbox{b($^{\circ}$)} &  $V$ & $J$ & $H$ & $K_{\rm s}$ &   $J_{\rm VVV}$ 
& {\rm $H_{\rm VVV}$} & {\rm $K_{\rm s VVV}$} & E($B-V$)$^{a}$ & E($B-V$)$^{b}$ & \hbox{S/N}\\
\noalign{\vskip 0.2cm}
\noalign{\hrule\vskip 0.2cm}
\noalign{\vskip 0.2cm}
221 & 18033285-3117421 & 18:03:32.80 & $-$31:17:42.04 & 359.92 & $-$4.54 & 17.9 & 13.95 & 13.08 & 12.84 & 13.85 & 13.08 & 12.84 & 0.73 & 0.85 & 101 \\
224 & 18034522-3117379 & 18:03:45.18 & $-$31:17:37.72 & 359.94 & $-$4.57 & 17.9 & 13.96 & 13.10 & 12.84 & 13.77 & 13.02 & 12.79 & 0.77 & 0.90 &  79 \\
230 & 18033933-3114044 & 18:03:39.28 & $-$31:14:04.24 & 359.98 & $-$4.53 & 18.6 & 14.70 & 13.94 & 13.56 & 14.62 & 13.85 & 13.61 & 0.76 & 0.89 &  65 \\
235 & 18032741-3109441 & 18:03:27.40 & $-$31:09:43.96 &   0.02 & $-$4.45 & 16.2 & 12.63 & 11.81 & 11.60 & 12.58 & 11.71 & 11.57 & 0.72 & 0.83 & 175 \\
238 & 18031238-3106210 & 18:03:12.35 & $-$31:06:20.92 &   0.05 & $-$4.38 & 17.0 & 13.14 & 12.24 & 12.02 & 13.03 & 12.25 & 11.95 & 0.71 & 0.82 & 152 \\
\noalign{\smallskip} \hline 
\end{tabular}
\tablebib{$^{a}$Schlafly \& Finkbeiner (2011), $^{b}$Schlegel et al. (1998)}
\end{flushleft}
\label{starmag}
\end{table*}

\begin{table*}
\caption{Photometric temperatures derived using the calibrations by Alonso et al. (1999) 
for several colours and the final temperature adopted. 
Colours from 2MASS and VVV catalogues were used, with reddening E($B-V$) 
based on Schlegel et al. (1998) (first line) and Schlafly \& Finkbeiner (2011) 
(second line) for each star.}             
\label{temp}      
\centering                          
\begin{tabular}{crcccccc}        
\hline\hline                 
\noalign{\smallskip}
\hbox{Star} & \hbox{\Teff($V-K$)} & \hbox{\Teff($V-K$)} & \hbox{\Teff($J-H$)} & \hbox{\Teff($J-H$)} & \hbox{\Teff($J-K$)} & \hbox{\Teff($J-K$)} & \hbox{Average \Teff}\\
\hbox{} & \hbox{2MASS} & \hbox{VVV} & \hbox{2MASS} & \hbox{VVV} & \hbox{2MASS} & \hbox{VVV} & \hbox{(K)} \\
\noalign{\smallskip}
\hline
\noalign{\smallskip}
\hbox{221} & 4179.3 & 4172.5 & 4281.9 & 4304.9 & 4404.8 & 4478.9 & 4303.7\\
\hbox{}    & 4401.3 & 4392.8 & 4400.9 & 4425.4 & 4576.7 & 4658.7 & 4476.0\\
\hbox{224} & 4249.3 & 4211.3 & 4348.7 & 4415.8 & 4436.3 & 4644.4 & 4384.3\\
\hbox{}    & 4502.9 & 4455.1 & 4479.0 & 4550.9 & 4621.7 & 4853.9 & 4577.3\\
\hbox{230} & 4244.7 & 4267.8 & 4666.3 & 4348.2 & 4379.3 & 4526.9 & 4405.5\\
\hbox{}    & 4493.2 & 4522.1 & 4817.5 & 4476.7 & 4555.6 & 4719.7 & 4597.5\\
\hbox{235} & 4476.1 & 4443.6 & 4437.9 & 3997.6 & 4640.7 & 4445.2 & 4406.9\\
\hbox{}    & 4764.5 & 4723.9 & 4564.5 & 4096.8 & 4834.1 & 4617.7 & 4600.3\\
\hbox{238} & 4182.7 & 4135.7 & 4180.5 & 4258.3 & 4336.5 & 4261.4 & 4225.9\\
\hbox{}    & 4396.8 & 4338.9 & 4289.0 & 4371.5 & 4495.3 & 4412.7 & 4384.0\\
\noalign{\smallskip}
\hline
\end{tabular}
\end{table*}

\section {Observations and reductions}

We used the UVES spectrograph (Dekker et al. 2000), in FLAMES-UVES mode, for the observation 
of five metal-poor ([Fe/H]~$\sim-1$) bulge stars, at a high resolution of R~=~45, 000 with a 
slit width of 0.8''. Centring the wavelength at 5800~{\rm \AA}, the spectral wavelength range 
4800~$-$~6800~{\rm \AA} with a gap at 5708~$-$~5825~{\rm \AA} was obtained. The red chip 
(5800~$-$~6800~{\rm \AA}) has ESO CCD\#20, an MIT backside illuminated, with 4096~x~2048 pixels, 
and pixel size 15~x~15~$\mu$m. The blue chip (4800~$-$~5800~{\rm \AA}) uses ESO Marlene EEV CCD\#44, 
backside illuminated, with 4102~x~2048 pixels, and pixel size 15~x~15~$\mu$m. The pixel scale is 
0.0147~{\rm \AA}/pix, with $\sim7.5$ pixels per resolution element at 6000~{\rm \AA}.

\begin{figure}
\centering
\includegraphics[width=\hsize]{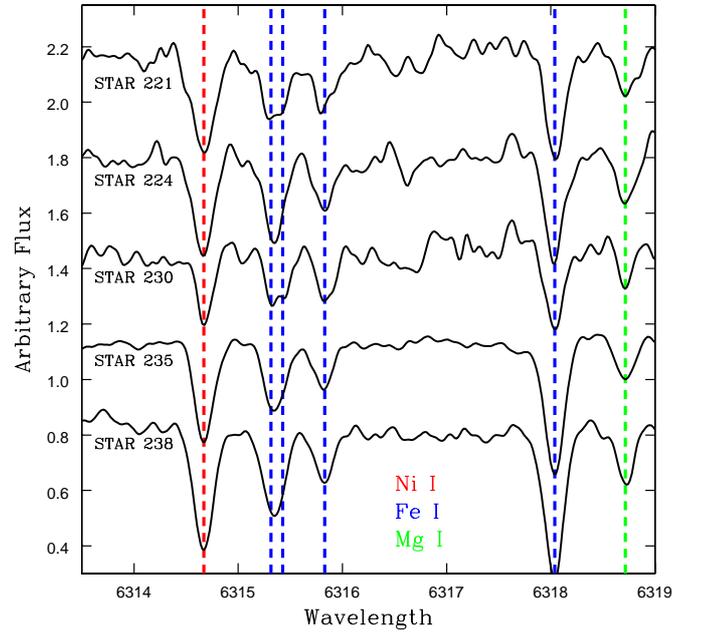}
\caption{Portion of the final data. The dashed lines show \ion{Ni}{I} (red), 
\ion{Fe}{I} (blue), and \ion{Mg}{I} (green) lines located in this wavelength range.}
\label{espectros}
\end{figure}

The log of observations is given in Table \ref{log}. The data were reduced using the UVES 
pipeline, within ESO/Reflex software (Ballester et al. 2000; Modigliani et al. 2004). 
The spectra were flatfielded, optimally-extracted and wavelength calibrated with the FLAMES-UVES 
pipeline. The spectra were normalized, corrected for radial-velocity shift, and combined to 
produce the final average data. Figure \ref{espectros} ilustrates the quality of the spectra 
for the five sample stars.

\subsection{Radial velocities}

In Table \ref{vradial}, we report the geocentric and heliocentric radial velocities measured 
with IRAF/FXCOR for each of the 10 runs, together with their mean values. A solar spectrum 
was adopted as the template. We used a solar synthetic spectrum to confirm the correction for 
the radial-velocity shift. We note that the heliocentric radial velocity for the star 221 obtained 
in run 4 is excessively different in comparison to others, so it was excluded from the final 
average spectrum. In addition, the IRAF routine was not able to measure the radial velocities 
for the star 230 using two runs (4 and 5), and they were also discarded. 


\section{Photometric stellar parameters}

\subsection{Temperatures}

The selected stars, their OGLE and 2MASS designations, coordinates, 
magnitudes, and S/N values corresponding to an average of clean windows 
in the range 6400-6500~{\AA}, are given in Table~\ref{starmag}.
$V$ magnitudes computed using individual reddening from Schlegel et al. (1998) 
in the direction of each star are adopted from the ARGOS survey 
(Freeman et al. 2013), $JHK_s$ magnitudes from 2MASS (Skrutskie et al. 2006), 
and VVV surveys (Saito et al. 2012). In this section we derive the photometric 
temperatures, to compare with results from the ARGOS survey, as explained below.

We calculated photometric temperatures based on three colours: ($V-K$), ($J-H$), and ($J-K$). 
Calibrations by Alonso et al. (1999) were applied, with reddening E($B-V$) computed with the 
Galactic reddening and extinction calculator from the Infrared Processing and Analysis Center 
(IRSA)\footnote{http://irsa.ipac.caltech.edu/applications/DUST/}. Results based on 
Schlegel et al. (1998) and on Schlafly \& Finkbeiner (2011) were used. The extinction laws 
given by Rieke \& Lebofsky (1985) were adopted. Colours from 2MASS were transformed into 
the ESO photometric system and from this into the TCS (Telescopio Carlos S\'anchez) system, 
following the relations established by Carpenter (2001) and Alonso et al. (1998). 
The VVV $JHK_{s}$ colours were transformed to the 2MASS $JHK_{s}$ system, using relations by 
Soto et al. (2013). 

The derived photometric effective temperatures are listed in Table~\ref{temp}. In our sample, 
the values obtained using the reddening from Schlafly \& Finkbeiner (2011) are 
$\Delta$\Teff~$=182\pm13$~K higher than the results with reddening from Schlegel et al. (1998). 
The more affected temperature is the one obtained with the colour ($V-K$), for which the average 
difference is $\Delta$\Teff~$=243\pm22$~K. The differences between temperatures derived with 
the VVV and 2MASS $JHK_{s}$ colours are $\Delta$\Teff~$=12\pm45$~K, which indicates that there 
is no significant trend in the temperature owing to the survey chosen for our sample. 
The outlier is the star 235: the temperatures obtained with 2MASS $JHK_{s}$ colours are 
$\Delta$\Teff~$=232\pm9$~K higher than the results from the VVV survey.

Table \ref{paramargo} shows the parameters obtained from the ARGOS survey to our set of stars. 
As described in Freeman et al. (2013), the effective temperatures were derived from the ($J-K$) 
colours using the calibration from Bessell et al. (1998) with interstellar reddening from 
Schlegel et al. (1998). The values for \logg, [Fe/H], and [$\alpha$/Fe] were determined by 
comparing the observed spectra with a grid of synthetic spectra computed in LTE with the 
code MOOG 2010 (Sneden 1973) and using 1D model atmospheres, described in 
Castelli \& Kurucz (2004). A constant microturbulence velocity $\xi=2.0$~\kms~was adopted in 
their method. 

The photometric temperatures derived in this work are systematically lower than the 
results from ARGOS: $\Delta$\Teff$=388\pm54$~K if the reddening from Schlegel et al. (1998) 
is adopted and $\Delta$\Teff$=206\pm59$~K for the reddening maps of Schlafly \& Finkbeiner (2011). 
These differences are smaller when our temperatures derived with the ($J-K$) colours are compared: 
$\Delta$\Teff$=278\pm58$~K and $\Delta$\Teff$=98\pm66$~K, respectively. Indeed, 
Freeman et al. (2013) report a mean temperature lower by 100~K when the empirical calibration 
from Alonso et al. (1999) is applied, and this difference is higher (up to 200~K) for the 
most metal-poor stars in the sample.

\begin{table}
\caption{Galactocentric velocities, atmospheric parameters, and enhancement 
in $\alpha-$elements for the present sample from the Galactic bulge ARGOS Survey.}             
\label{paramargo}      
\centering                          
\begin{tabular}{crcccc}        
\hline\hline                 
\noalign{\smallskip}
\hbox{Star} & \hbox{Vgal} & \hbox{\Teff} & \hbox{\logg} & \hbox{[Fe/H]} & \hbox{[$\alpha$/Fe]}   \\
\hbox{} & \hbox{(\kms)} & \hbox{(K)} & \hbox{[cgs]} & \hbox{} & \hbox{}   \\
\noalign{\smallskip}
\hline
\noalign{\smallskip}
\hbox{221} & $-$96.06 & 4669.54 & 1.6 & $-$0.80 & 0.40 \\
\hbox{224} &$-$105.17 & 4757.18 & 1.8 & $-$0.82 & 0.35 \\
\hbox{230} & $-$71.73 & 4697.27 & 1.4 & $-$0.84 & 0.03 \\
\hbox{235} &   146.01 & 4929.72 & 2.4 & $-$0.80 & 0.52 \\
\hbox{238} &$-$136.31 & 4613.03 & 2.4 & $-$0.80 & 0.33 \\
\noalign{\smallskip}
\hline
\end{tabular}
\end{table}

\section{Spectroscopic stellar parameters}

\subsection{Equivalent widths}

To derive the atmospheric parameters, we measured the equivalent 
widths (EW) of selected iron lines, using the IRAF software. 
We decided to retain the lines with $10<$~EW~$<100$~m{\AA} located in 
the range 6100~$-$~6800~{\AA} to derive the atmospheric parameters, 
which is the spectral region with the highest S/N ratio available. 
The EW values measured manually were adopted since this method allows 
a better continuum placement and an individual evaluation of each line.

Table. \ref{EW_measurements} presents the complete list of lines, 
describing the atomic data, the EW measured with IRAF, 
and the individual iron abundance derived using the atmospheric 
parameters adopted.

\begin{table*}
\caption{Spectroscopic parameters adopted for each star. We also present the number of iron lines 
used for each star to derive the atmospheric parameters. 
The difference $\Delta_{II-I}=[\ion{Fe}{II}/\hbox{H}]~-~[\ion{Fe}{I}/\hbox{H}]$ shows the quality in the ionization 
equilibrium, and the parameters from the linear fitting $[\ion{Fe}{I}/\hbox{H}]=a_{EW}*log(EW/\lambda)+b_{EW}$ 
for microturbulence and $[\ion{Fe}{I}/\hbox{H}]=a_{exc}*exc.pot+b_{exc}$ for effective temperature.}             
\label{param}      
\scalefont{0.75}
\centering                          
\begin{tabular}{crcccccrcrrrrr}        
\hline\hline                 
\noalign{\smallskip}
\hbox{Star} & \hbox{\Teff} & \hbox{\logg} & \hbox{[FeI/H]} & \hbox{[FeII/H]} & \hbox{[Fe/H]$_{model}$} & \hbox{$\xi$} & 
\hbox{\#\ion{Fe}{I}} & \hbox{\#\ion{Fe}{II}} & \hbox{$\Delta_{II-I}$} 
& \hbox{a$_{EW}$} & \hbox{b$_{EW}$} & \hbox{a$_{exc}$} & \hbox{b$_{exc}$}\\
\hbox{} & \hbox{(K)} & \hbox{[cgs]} & \hbox{} & \hbox{} & \hbox{} & \hbox{(\kms)} & \hbox{} & \hbox{} & \hbox{}
& \hbox{} & \hbox{} & \hbox{} & \hbox{} \\
\noalign{\smallskip}
\hline
\noalign{\smallskip}
\hbox{221} & 4620 & 2.0 & $-$0.88$\pm$0.24 & $-$0.91$\pm$0.17 & $-$0.90 & 1.0 & 31 & 2 & $-$0.035  & $+$0.00$\pm$0.12 & $+$6.63$\pm$0.55 & $+$0.004$\pm$0.045 & $+$6.61$\pm$0.15 \\
\hbox{224} & 5000 & 3.5 & $-$0.62$\pm$0.36 & $-$0.60$\pm$0.30 & $-$0.65 & 0.8 & 61 & 4 & $+$0.015  & $+$0.02$\pm$0.10 & $+$6.96$\pm$0.48 & $+$0.013$\pm$0.044 & $+$6.84$\pm$0.15 \\
\hbox{230} & 4960 & 3.0 & $-$1.04$\pm$0.40 & $-$1.03$\pm$0.18 & $-$1.10 & 0.8 & 58 & 5 & $+$0.0038 & $+$0.02$\pm$0.11 & $+$6.57$\pm$0.52 & $+$0.087$\pm$0.056 & $+$6.17$\pm$0.19 \\
\hbox{235} & 4680 & 2.2 & $-$0.94$\pm$0.15 & $-$0.95$\pm$0.09 & $-$0.95 & 1.1 & 45 & 7 & $-$0.012  & $+$0.00$\pm$0.03 & $+$6.59$\pm$0.17 & $-$0.002$\pm$0.024 & $+$6.57$\pm$0.08 \\
\hbox{238} & 4720 & 2.9 & $-$0.43$\pm$0.18 & $-$0.42$\pm$0.07 & $-$0.50 & 1.0 & 33 & 5 & $-$0.0077 & $+$0.00$\pm$0.07 & $+$7.07$\pm$0.35 & $+$0.027$\pm$0.031 & $+$6.98$\pm$0.11 \\
\noalign{\smallskip}
\hline
\end{tabular}
\end{table*}

\subsection{Atmospheric parameters}

The photometric temperatures, together with the gravity and 
metallicity values from the ARGOS survey as given in Table \ref{paramargo},
are adopted as a first guess to calculate the excitation 
and ionization equilibria of \ion{Fe}{I} and \ion{Fe}{II} lines. 
The MARCS spherical model atmosphere grids (Gustafsson et al. 2008) with 1~M$_{\odot}$ 
and the code Turbospectrum (Alvarez \& Plez 1998) in the equivalent width mode 
were used, with solar abundances adopted from Asplund et al. (2009). 
Applying an automatic routine on a grid of models with $\Delta$\Teff~$=20$~K, 
$\Delta$\logg~$=0.1$~[cgs], and $\Delta\xi=0.1$~\kms, the final surface 
gravity \logg~was chosen to minimize $[\ion{Fe}{II}/\hbox{H}]~-~[\ion{Fe}{I}/\hbox{H}]$, 
the final microturbulence velocity $\xi$ was chosen to minimize the dependence of 
[\ion{Fe}{I}/H] on log~(EW/$\lambda$), and the final temperature was obtained by 
the excitation equilibrium. The grid was recomputed successively with a new metallicity 
in each step, and the range used for each parameter was selected to avoid local solutions. 

\begin{figure}
\centering
\resizebox{90mm}{!}{\includegraphics[angle=0]{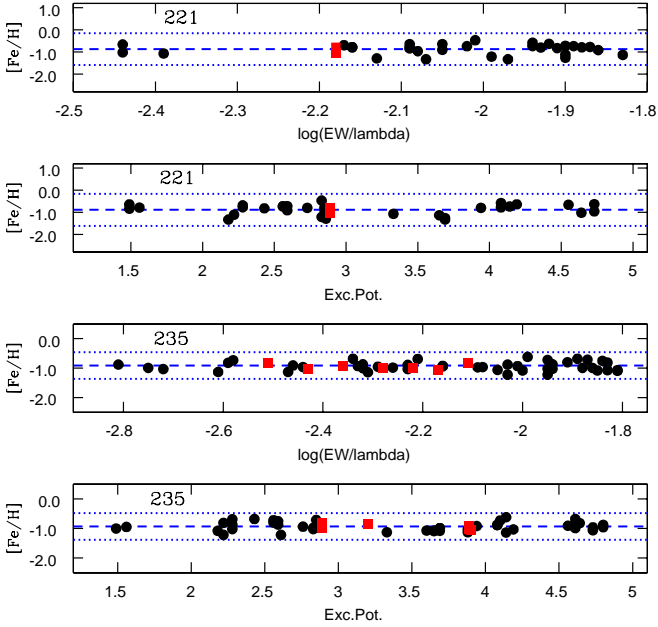}}
\caption{Ionization and excitation equilibria of Fe lines for the stars 221 and 235, 
using the newly derived atmospheric parameters. The black dots are the abundances 
obtained from the \ion{Fe}{I} lines, the red squares are those from the \ion{Fe}{II} lines, 
the blue dashed lines represent the linear fit to data, and the dotted blue lines 
are the same linear fit moved vertically by $\pm3\sigma$, where $\sigma$ is the 
standard error of the mean.}
\label{221_235}
\end{figure}

Figure \ref{221_235} shows the excitation and ionization equilibria for two different 
typical cases: the star 221, which presents a spectrum with a low S/N; and the star 235, 
which presents a high-quality spectrum. Lines with abundances out of the region limited by 
$\pm3\sigma$, where $\sigma$ is the standard error of the mean, were removed from the computations 
of final metallicities.

The black dots are the 
abundances obtained from \ion{Fe}{I} lines, and the red squares are the results 
from the \ion{Fe}{II} lines. The blue dashed lines represent the linear fit to data, 
and the blue dotted lines are the same function moved vertically by . 

The derived stellar parameters are reported in Table \ref{param}, as well as the number 
of \ion{Fe}{I} and \ion{Fe}{II} lines retained, the difference 
$\Delta_{II-I}=[\ion{Fe}{II}/\hbox{H}]~-~[\ion{Fe}{I}/\hbox{H}]$ obtained with the final models, 
and the parameters from the linear fit to data in each case. The angular coefficients
and the values of $\Delta_{II-I}$ are null within the error bar, which indicate that there 
are no relevant trends in the excitation and ionization equilibria. 
The parameters can be compared with the results derived from the mid-resolution survey ARGOS, 
reported in Table \ref{paramargo}.

\section{Abundance ratios}

\begin{table*}
\caption{Mean LTE abundances of the elements derived in the present work.}             
\label{finalAbund}      
\scalefont{1.0}
\centering                          
\begin{tabular}{crrrrrrrrrrr}        
\hline\hline                 
\noalign{\smallskip}
\hbox{Element} & \hbox{A(X)$_{\odot}$} & \hbox{A(X)} & \hbox{[X/Fe]} & \hbox{A(X)} & \hbox{[X/Fe]} & \hbox{A(X)} & \hbox{[X/Fe]} & \hbox{A(X)} & \hbox{[X/Fe]} & \hbox{A(X)} & \hbox{[X/Fe]} \\
\noalign{\smallskip}
\hline
\noalign{\smallskip}
& & \multicolumn{2}{c}{221} & \multicolumn{2}{c}{224} & \multicolumn{2}{c}{230} & \multicolumn{2}{c}{235} & \multicolumn{2}{c}{238}  \\
\noalign{\smallskip}
\hline
\noalign{\smallskip}
\hbox{\ion{Fe}{I}*}  & $+$7.50 & $+$6.62 & $-$0.88 & $+$6.88 & $-$0.62 & $+$6.46 & $-$1.04 & $+$6.56 & $-$0.94 & $+$7.07 & $-$0.43 \\
\hbox{\ion{Fe}{II}*} & $+$7.50 & $+$6.59 & $-$0.91 & $+$6.90 & $-$0.60 & $+$6.47 & $-$1.03 & $+$6.55 & $-$0.95 & $+$7.08 & $-$0.43 \\
\hbox{\ion{O}{I}}    & $+$8.69 & $+$8.40 & $+$0.61 & ------- & ------- & $+$8.25 & $+$0.60 & $+$8.40 & $+$0.66 & $+$8.85 & $+$0.59 \\
\hbox{\ion{Na}{I}}   & $+$6.24 & $+$5.68 & $+$0.33 & $+$5.58 & $-$0.06 & $+$4.90 & $-$0.31 & $+$5.21 & $-$0.09 & $+$5.93 & $+$0.12 \\
\hbox{\ion{Mg}{I}}   & $+$7.60 & $+$7.22 & $+$0.52 & $+$7.50 & $+$0.51 & $+$7.19 & $+$0.63 & $+$7.13 & $+$0.47 & $+$7.45 & $+$0.28 \\
\hbox{\ion{Al}{I}}   & $+$6.45 & $+$6.00 & $+$0.45 & $+$6.29 & $+$0.45 & $+$5.77 & $+$0.36 & $+$5.79 & $+$0.28 & $+$6.23 & $+$0.21 \\
\hbox{\ion{Si}{I}}   & $+$7.51 & $+$6.92 & $+$0.31 & $+$7.25 & $+$0.35 & $+$6.98 & $+$0.50 & $+$6.78 & $+$0.22 & $+$7.23 & $+$0.14 \\
\hbox{\ion{Ca}{I}}   & $+$6.34 & $+$5.67 & $+$0.23 & $+$6.01 & $+$0.28 & $+$5.46 & $+$0.16 & $+$5.75 & $+$0.36 & $+$6.10 & $+$0.18 \\
\hbox{\ion{Ti}{I}}   & $+$4.95 & $+$4.38 & $+$0.33 & $+$4.71 & $+$0.37 & $+$4.28 & $+$0.37 & $+$4.39 & $+$0.39 & $+$4.87 & $+$0.34 \\
\hbox{\ion{Ti}{II}}  & $+$4.95 & $+$4.51 & $+$0.46 & $+$4.80 & $+$0.46 & $+$4.34 & $+$0.42 & $+$4.38 & $+$0.38 & $+$4.82 & $+$0.30 \\
\hbox{\ion{Y}{I}}    & $+$2.21 & $+$1.60 & $+$0.29 & $+$1.95 & $+$0.12 & $+$1.90 & $+$0.73 & $+$1.60 & $+$0.34 & $+$1.92 & $+$0.14 \\
\hbox{\ion{Zr}{I}}   & $+$2.56 & $+$2.40 & $+$0.72 & $+$2.60 & $+$0.63 & $+$2.40 & $+$0.86 & $+$2.33 & $+$0.69 & $+$2.60 & $+$0.45 \\
\hbox{\ion{Ba}{II}}  & $+$2.18 & $+$1.60 & $+$0.32 & $+$1.80 & $+$0.23 & $+$1.63 & $+$0.48 & $+$1.90 & $+$0.67 & $+$2.10 & $+$0.35 \\
\hbox{\ion{La}{II}}  & $+$1.10 & $+$0.30 & $+$0.10 & $+$1.03 & $+$0.54 & ------- & ------- & $+$0.68 & $+$0.53 & $+$0.90 & $+$0.23 \\
\hbox{\ion{Eu}{II}}  & $+$0.52 & $+$0.00 & $+$0.38 & $+$0.35 & $+$0.44 & $-$0.10 & $+$0.42 & $+$0.05 & $+$0.48 & $+$0.56 & $+$0.47 \\
\noalign{\smallskip}
\hline
\end{tabular}
\tablefoot{ *: [X/H] is used in place of [X/Fe].}
\end{table*}

A line-by-line fitting was carried out to derive the abundances, 
using the spectrum synthesis code Turbospectrum (Alvarez \& Plez 1998), 
which includes scattering 
in the blue and UV domain, molecular dissociative equilibrium, and collisional 
broadening by H, He, and H$_{2}$, following Anstee \& O'Mara (1995), 
Barklem \& O'Mara (1997), and Barklem et al. (1998). The atomic line lists 
were adopted from the Vienna Atomic Line Database compilation (VALD3; Piskunov et al. 1995), 
together with the Turbospectrum molecular line lists (B. Plez, private communication). 
For lines used to derive abundances, as reported in Table \ref{linelist}, 
the oscillator strengths were adopted from Barbuy et al. (2014), except 
when described. Hyperfine structures were adopted for the lines relevantly 
affected by this effect. Tables \ref{finalAbund} and \ref{CN} show the adopted final 
abundances.

\subsection{Carbon and nitrogen}

To evaluate the adopted line list in the regions selected for 
carbon and nitrogen abundances, we used the Arcturus spectrum (Hinkle et al. 2000) 
as a reference star. Our benchmark analysis is based on the stellar parameters 
described in Mel\'endez et al. (2003): \Teff~$=4275$~K, \logg~$=1.55$~[cgs], 
[Fe/H]~$=-0.54$, and $\xi=1.65$~\kms. We adopted chemical abundances from 
Ram\'irez \& Allende Prieto (2011) and Mel\'endez et al. (2003), 
as presented in Table \ref{arcturus_abund}.

\begin{table}
\caption{Adopted Arcturus abundaces.}   
\label{arcturus_abund} 
\centering                  
\begin{tabular}{c c c c} 
\hline\hline             
El. & A(X)$_{Arcturus}$ & El. & A(X)$_{Arcturus}$ \\ 
\hline                  
\noalign{\smallskip}
C  & 8.32$^{[1]}$ & Ca & 5.94$^{[1]}$ \\   
N  & 7.68$^{[2]}$ & Sc & 2.81$^{[1]}$ \\   
O  & 8.66$^{[2]}$ & Ti & 4.66$^{[1]}$ \\   
Na & 5.82$^{[1]}$ & V  & 3.58$^{[1]}$ \\   
Mg & 7.47$^{[1]}$ & Cr & 4.99$^{[1]}$ \\   
Al & 6.26$^{[1]}$ & Mn & 4.74$^{[1]}$ \\   
Si & 7.30$^{[1]}$ & Co & 4.71$^{[1]}$ \\   
K  & 4.99$^{[1]}$ & Ni & 5.73$^{[2]}$ \\   
\hline                          
\end{tabular}
\tablebib{[1]: Ram\'irez \& Allende Prieto (2011); 
[2]: Mel\'endez et al. (2003).}
\end{table}

\begin{figure}
\centering
\resizebox{90mm}{!}{\includegraphics[angle=0]{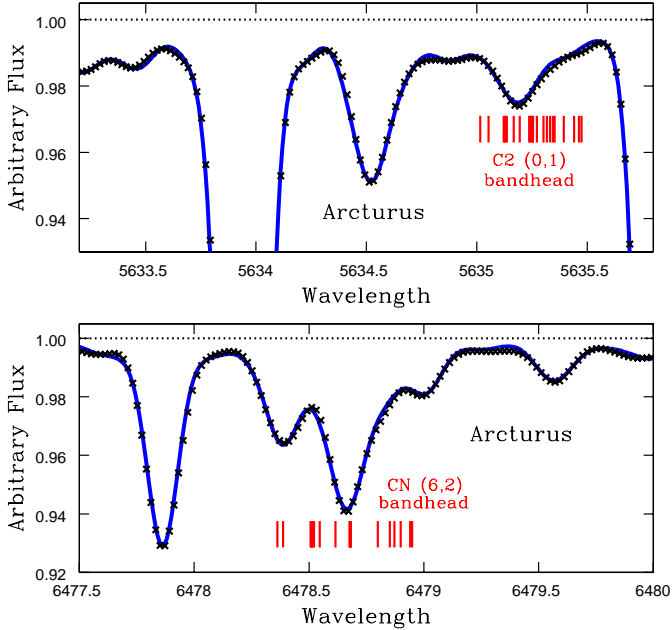}}
\caption{\textbf{Upper panel:} Fit to the C$_{2}$(1,0) molecular 
bandhead at 5635.3~{\AA} in Arcturus. Observations (black crosses) 
are compared with synthetic spectra computed with the adopted abundances 
(blue solid lines) from literature. Red marks show the positions of molecular lines. 
\textbf{Lower panel:} Fit to the CN(6,2) molecular bandhead at 
6478.5~{\AA} in Arcturus. Symbols are the same as in the upper panel.}
\label{arcturus}
\end{figure}

\begin{figure}
\centering
\resizebox{90mm}{!}{\includegraphics[angle=0]{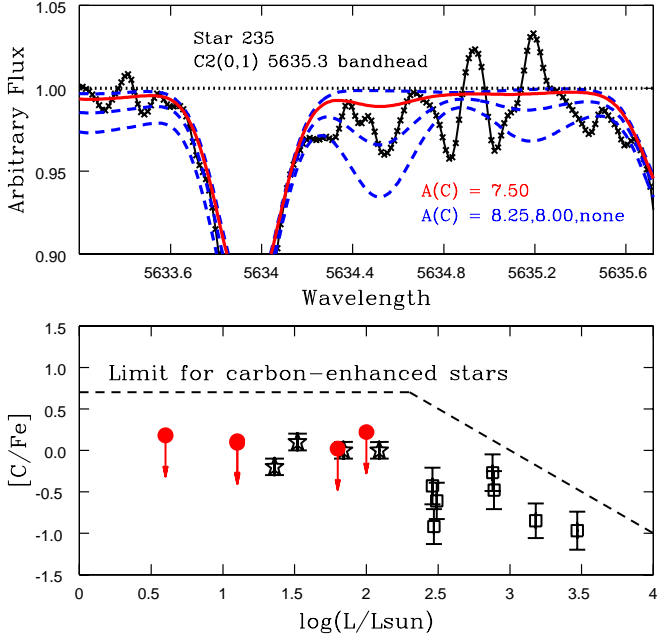}}
\caption{\textbf{Upper panel:} Fit to the C$_{2}$(1,0) bandhead 
at 5635.3~{\AA} in star 235. Observations (black crosses) are 
compared with synthetic spectra computed with the different abundances 
indicated in the figure (blue dashed lines), as well as with the adopted 
abundance (red solid lines), also indicated. 
\textbf{Lower panel:} comparison of the [C/Fe] abundance ratios derived 
in the present sample (filled red circles) with stars from the 
bulge blobular clusters M~62 (Yong et al. 2014, open black squares) 
and NGC~6522 (Barbuy et al. 2014, open black stars). The dashed black 
line corresponds to the limit for carbon-enhanced stars, as defined 
by Aoki et al. (2007).}
\label{carbon}
\end{figure}

To measure the carbon abundances we used the C$_{2}(0,1)$ molecular bandhead. 
The region is extended and a mean abundance was derived from the 
overall fit, however the bandhead at 5635.3~{\rm \AA} received 
more weight in the fitting procedure. The line list 
for $^{12}$C$_{2}$, $^{13}$C$_{2}$, and $^{12}$C$^{13}$C was 
adopted from Wahlin \& Plez (2005), which contains transitions from the 
Swan (d$^{3}\Pi$~-~a$^{3}\Pi$) electronic band. The solar isotopic 
fraction for $^{12}$C (98.9\%) and $^{13}$C (1.1\%) was adopted (Asplund et al. 2009). 
Figure \ref{arcturus} shows in the upper panel the synthetic spectrum 
computed for Arcturus (blue solid line), which is in very good agreement with 
observations. For the sample stars, the C$_{2}(0,1)$ molecular bandhead is located 
in the region observed with the blue chip, showing a lower S/N. 
An example can be seen in the fit to  star 235 shown in Fig. \ref{carbon} 
(upper panel). The C abundances were adopted as upper limits.

The derived abundances are presented in Table \ref{CN}. Beers \& Christlieb (2005) 
defined carbon-enhanced metal-poor stars (CEMP) as having [C/Fe]~$>+1.0$, 
but Aoki et al. (2007) presents a new definition, which takes into account the 
mixing events in evolved stars and the consequently lower carbon abundance on their 
surface. Following Aoki et al. (2005, 2007), we assumed the mass of the stars 
to be $0.8$~M$_\odot$ to calculate the luminosities 
$L/L_{\odot}\propto(M/M_{\odot})(g/g_{\odot})^{-1}(T_{eff}/T_{eff\odot})^{4}$, 
and in Fig. \ref{carbon} (lower panel) we show the [C/Fe] abundance ratios 
as a function of the luminosity log($L/L_{\odot}$) for our sample. The limits for 
CEMP stars are also presented, showing that our sample consists of carbon-normal 
metal-poor stars (non-CEMP). 

For comparison, we included carbon abundances of bulge stars from the literature. 
The open black squares represent seven stars in the globular cluster M~62 (NGC~6266) 
studied in high-resolution by Yong et al. (2014). This object is located at J(2000) 
$\alpha=17^{h}01^{m}12.60^{s}$ and $\delta=-30^{\circ}06'44.5''$ (Di Criscienzo et al. 2006), 
or $l=353.5746^{\circ}$ and $b=+7.3196^{\circ}$, therefore projected in the bulge. 
The open black stars represent the results of the high-resolution abundance analysis from 
Barbuy et al. (2014) for the globular cluster NGC~6522, which is located at J(2000) 
$\alpha=18^{h}03^{m}34.08^{s}$ and $\delta=-30^{\circ}02'02.3''$, or $l=1.0246^{\circ}$ 
and $b=-3.9256^{\circ}$ (Barbuy et al. 2009), and therefore also projected in the bulge.

The nitrogen abundance was derived using the CN A$^{2}\Pi$~-~X$^{2}\Sigma$ 
red system, based on the CN$(6,2)$~6478.48~{\rm \AA} bandhead. The CN line list is a 
compilation by B. Plez (private communication), using data from Cerny et al. (1978), 
Kotlar et al. (1980), Larsson et al. (1983), Bauschlicher et al. (1988), 
Ito et al. (1988a, 1988b), Prasad \& Bernath (1992), Prasad et al. (1992), and 
Rehfuss et al. (1992). All four isotope combinations $^{12}$C$^{14}$N, $^{12}$C$^{15}$N, 
$^{13}$C$^{14}$N, and $^{13}$C$^{15}$N were treated with nitrogen solar isotopic fraction 
$^{14}$N (99.8\%) and $^{15}$N (0.2\%) from Asplund et al. (2009). 
The synthetic spectrum computed for Arcturus (blue solid line) in this 
region is shown in Fig. \ref{arcturus} (lower panel), showing good agreement with 
observations.

\begin{figure}
\centering
\resizebox{90mm}{!}{\includegraphics[angle=0]{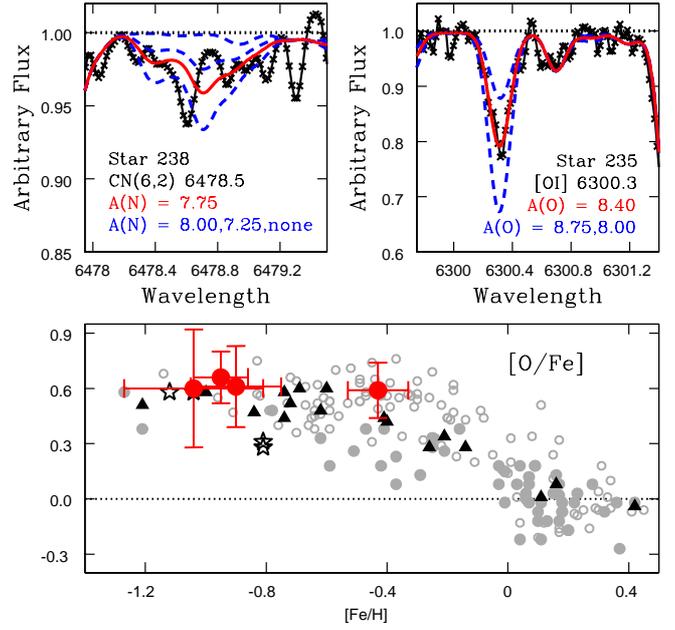}}
\caption{\textbf{Upper left panel:} Fit to the 
CN$(6,2)$~6478.48~{\rm \AA} bandhead in star 238. 
\textbf{Upper right panel:} Fit to the [\ion{O}{I}]~6300.3~{\AA} 
line in star 235. Symbols are the same as in Fig. \ref{carbon} 
(upper panel). \textbf{Lower panel:} [O/Fe] abundance 
ratio as a function of the metallicity for the five sample stars 
(filled red circles), compared with literature abundances from 
Bensby et al. (2013; filled black triangles), Barbuy et al. (2014; 
open black stars), Johnson et al. (2014; open grey circles), 
and Barbuy et al. (2015; filled grey circles).}
\label{N_O}
\end{figure}

For the sample stars, the selected molecular transitions are weak and the noise becomes 
more evident, as shown in Fig. \ref{N_O} (upper left panel) for star 238. Table \ref{CN} shows 
the derived N abundances which, owing to the previous discussion, must be used with caution. 
The difficulty in defining the local continuum does not permit us to determine the N abundance 
in the star 224. 

\begin{table}
\caption{Carbon and nitrogen abundances [X/Fe] from C$_{2}$ and 
CN bandheads.}
\label{CN}
\scalefont{0.9}
\centering                          
\begin{tabular}{ccccccc}        
\hline\hline
\noalign{\smallskip}
\hbox{Species} &  \hbox{$\lambda$({\rm\AA})} & \hbox{221} & \hbox{224} & \hbox{230} & \hbox{235} & \hbox{238}\\
\noalign{\smallskip}
\hline
\noalign{\smallskip}
\hbox{[C/Fe] C$_{2}(0,1)$} & 5635.3 & $<+$0.2 & $<+$0.2 & $<+$0.1 & $<$0.0 & $<+$0.1 \\
\hbox{[N/Fe] CN$(6,2)$}    & 6478.5 & $+$0.82 & ------- & $+$0.71 & $+$0.97 & $+$0.35 \\
\noalign{\smallskip}
\hline
\end{tabular}
\end{table}

\begin{figure}
\centering
\resizebox{90mm}{!}{\includegraphics[angle=0]{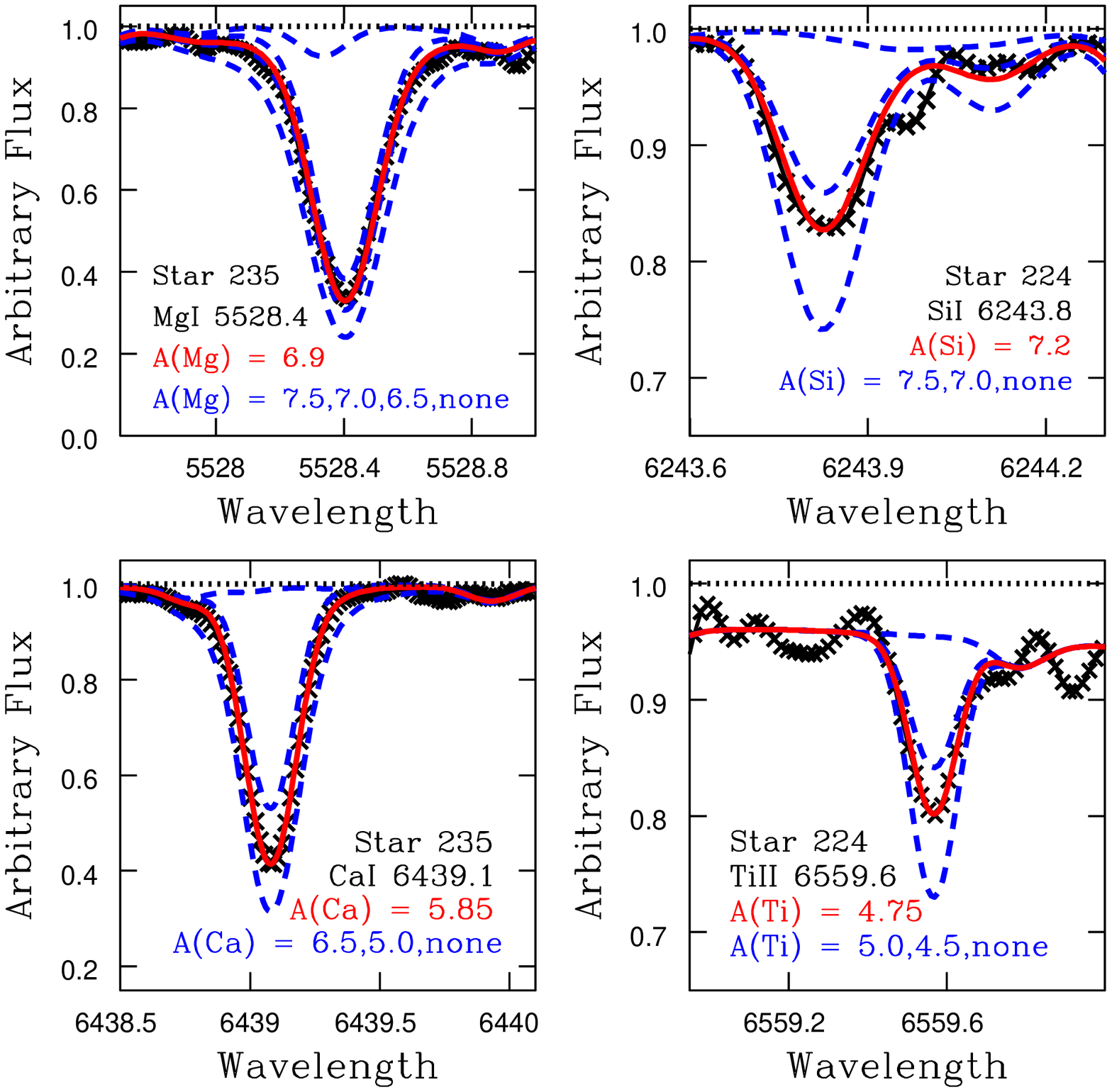}}
\caption{\textbf{Upper left panel:} Fit to the 
\ion{Mg}{I}~5528.4~{\AA} line in star 235. 
\textbf{Upper right panel:} Fit to the \ion{Si}{I}~6243.8~{\AA} 
line in star 224. 
\textbf{Lower left panel:} Fit to the \ion{Ca}{I}~6439.1~{\AA} 
line in star 235. \textbf{Lower right panel:} Fit to the 
\ion{Ti}{II}~6559.6~{\AA} line in star 224. 
Symbols are the same as in Fig. \ref{carbon} (upper panel).}
\label{Mg_Si_Ca_Ti}
\end{figure}

\begin{figure}
\centering
\resizebox{90mm}{!}{\includegraphics[angle=0]{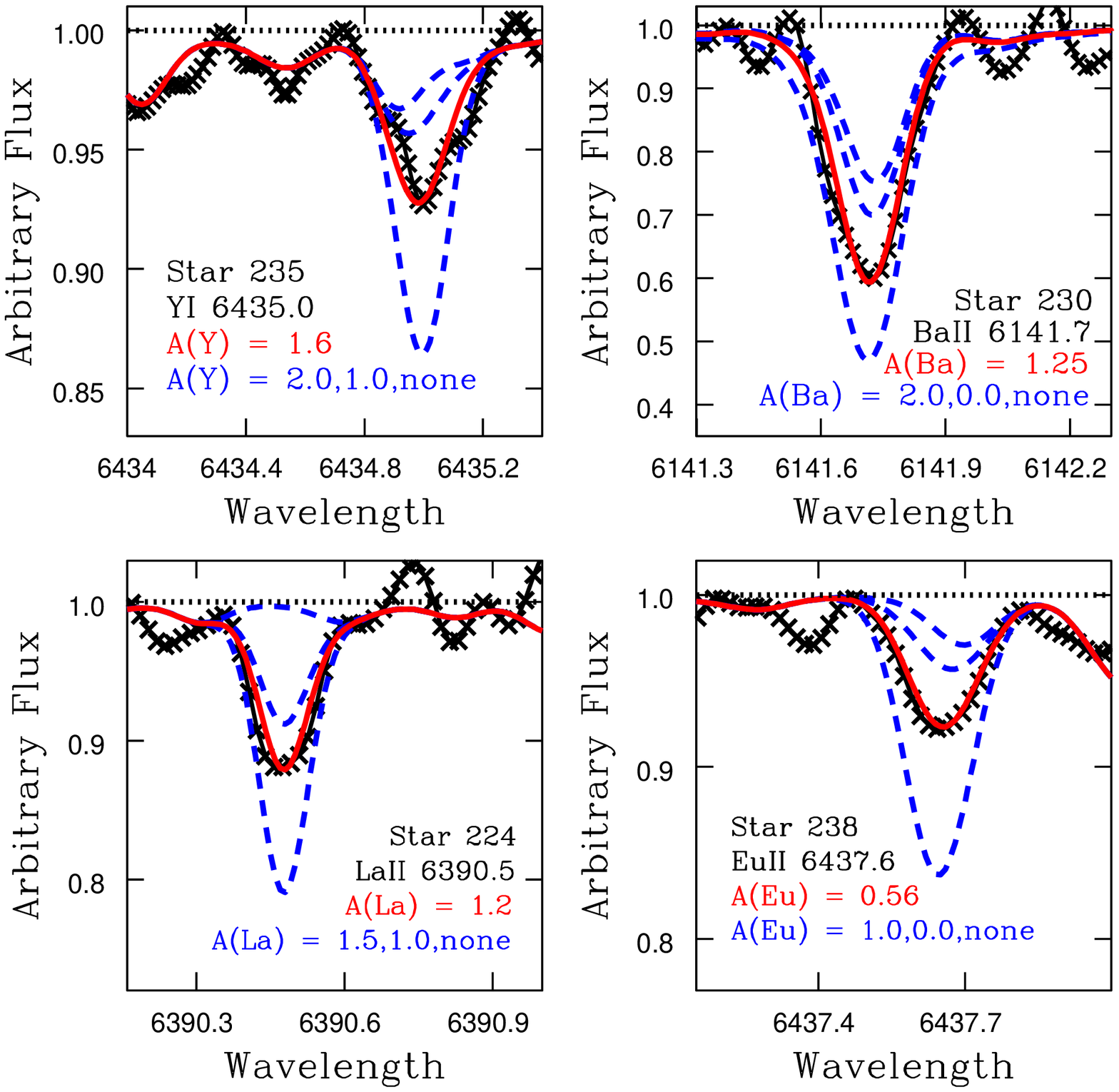}}
\caption{\textbf{Upper left panel:} Fit to the \ion{Y}{I}~6435.0~{\AA} 
line in star 235. \textbf{Upper right panel:} Fit to the 
\ion{Ba}{II}~6141.7~{\AA} line in star 230. 
\textbf{Lower left panel:} Fit to the \ion{La}{II}~6390.5~{\AA} 
line in star 224. \textbf{Lower right panel:} Fit to the 
\ion{Eu}{II}~6437.6~{\AA} line in star 238. 
Symbols are the same as in Fig. \ref{carbon} (upper panel).}
\label{Y_Ba_La_Eu}
\end{figure}

\subsection{Alpha elements}

The oxygen abundance was derived using the forbidden line [\ion{O}{I}]~6300.3~{\AA}, 
as shown in Fig. \ref{N_O} (upper right panel) for the star 235. 
We inspected the individual spectra, before combining, to check for possible blends 
with telluric lines, and we remove them from the final average when necessary. 
For star 224, the [\ion{O}{I}] line was strongly contaminated 
in all individual spectra, and consequently the oxygen abundance was not derived. 
Some individual spectra were also discarded owing to the higher noise level surrounding 
the [\ion{O}{I}] line, compared with the average, which allowed a better placement of the 
continuum. 

In Fig. \ref{N_O} (lower panel) we compare the [O/Fe] abundance ratios in the 
sample stars with the result in the bulge globular cluster NGC~6522 (Barbuy et al. 2014), 
with microlensed bulge dwarfs and subgiants stars from Bensby et al. (2013) selected to have 
ages older than 11 Gyr, with selected red giant branch stars in the Galactic bulge from Johnson 
et al. (2014), and with the giant stars from Barbuy et al. (2015). The solar oxygen abundance 
A(O)$_{\odot}=8.69$ (Asplund et al. 2009) adopted in our results is $0.08$~dex lower than 
A(O)$_{\odot}=8.77$ adopted in Barbuy et al. (2014, 2015) and, to ensure consistency among the 
abundance results, we shifted their values. This figure shows that our abundances are in agreement 
with previous results for bulge stars.

We checked four \ion{Mg}{I} lines located at 5528.4~{\AA}, 6318.7~{\AA}, 6319.24~{\AA}, and 6765.4~{\AA} 
to derive the magnesium abundance. The line at 5528.4~{\AA} is in the portion of the spectra with more noise, 
obtained with the blue chip, and was only useful in star 235, as shown in Fig. \ref{Mg_Si_Ca_Ti} (upper left panel). 
The silicon abundance was measured using ten \ion{Si}{I} lines, two of them (5665.5~{\AA} 
and 5690.4~{\AA}) located in the wavelengths measured with the blue chip, but with individual abundances that are 
consistent with results from the lines in the red portion of the spectra. In the upper right panel of 
Fig. \ref{Mg_Si_Ca_Ti}, we show the fit to the line at 6243.8~{\AA} in star 224.

The calcium abundance was derived after checking 19 \ion{Ca}{I} lines. The transition located at 
5601.3~{\AA} is the only line in the blue spectra and was used only for stars 235 and 238, giving individual 
abundances that are in agreement with the other lines. The result obtained from the \ion{Ca}{I}~6439.1~{\AA} 
line in star 235 is shown in Fig. \ref{Mg_Si_Ca_Ti} (lower left panel).

It was possible to inspect 14 \ion{Ti}{I} lines, with only the \ion{Ti}{I}~5689.5~{\AA} line located in the 
blue portion of the spectra. For the ionized species, six \ion{Ti}{II} lines were checked to obtain the 
titanium abundance, but three of them (5336.8~{\AA}, 5381.0~{\AA}, and 5418.7~{\AA}) are located in 
wavelengths of the blue portion of the spectra and they were only used for stars 235 and 238. In the lower right 
panel of Fig. \ref{Mg_Si_Ca_Ti}, we show the \ion{Ti}{II}~6559.6~{\AA} line measured in star 224. 
This line is located in the blue wing of the H$\alpha$ line, and it was necessary to take the hydrogen 
line in the spectrum synthesis into account.

\subsection{Odd-Z elements Na, Al}

The sodium abundances are based on four \ion{Na}{I} lines, 
located at 4982.8~{\AA}, 5688.2~{\AA}, 6154.2~{\AA}, 
and 6160.7~{\AA}. We did not use the resonance lines 
\ion{Na}{I}~5889.95~{\rm \AA} (D2) and \ion{Na}{I}~5895.92~{\rm \AA} 
(D1) because they are very sensitive to non-LTE effects. 
The only stable isotope 
$^{23}$Na has nuclear spin I~$=3/2$\footnote{Adopted from the 
Particle Data Group (PDG) collaboration: http://pdg.lbl.gov/} 
and therefore exhibits hyperfine structure (HFS). 
The hyperfine coupling constants are adopted from Das \& Natarajan (2008) 
and Marcassa et al. (1998). When not 
available in the literature, the hyperfine constants for a given level 
were assumed to be null. The line splitting was computed by employing 
a code made available by McWilliam et al. (2013). 

For aluminum, only \ion{Al}{I}~6696.0~{\AA} and \ion{Al}{I}~6698.7~{\AA} 
lines were available. The stable isotope $^{27}$Al has nuclear spin 
I~$=5/2$ and we adopted the hyperfine coupling constants from Nakai et al. (2007) 
and Belfrage et al. (1984) to compute the HFS. 

\subsection{Heavy elements}

We derive the abundances of the neutron-capture elements Y, Zr, Ba, La, 
and the reference r-element Eu.
We preferentially used lines of ionized species, since 
these elements are mostly in this form. 
 For strontium, we evaluated six \ion{Sr}{I} 
lines: 6408.5~{\AA}, 6504.0~{\AA}, 6546.8~{\AA}, 6550.2~{\AA}, 
6617.3~{\AA}, and 6791.0~{\AA}. All transitions are too weak for 
abundance purposes, consequently no [Sr/Fe] result is presented. 

The best yttrium line \ion{Y}{II}~6795.4~{\AA} is located in the border of 
the \'echelle spectrum and shows clearly fringes that prevent its use. 
The most reliable line from our spectra is \ion{Y}{II}~5544.6~{\AA}, 
which is located in the blue portion of the spectra and was not useful 
for the abundance determination. Consequently, the Y abundance in the sample 
is based on the \ion{Y}{I}~6435.0~{\AA} line, as shown in the upper left 
panel of Fig. \ref{Y_Ba_La_Eu} for star 235.

For zirconium, we checked three \ion{Zr}{II} lines located at 
5112.3~{\AA}, 5350.1~{\AA}, and 5350.3~{\AA}, but none is reliable 
for measuring abundances. Due to a lack of useful ionized lines, we measured 
abundances from three lines of \ion{Zr}{I}: 6127.47~{\rm \AA}, 6134.58~{\rm \AA}, 
and 6143.25~{\rm \AA}. Oscillator strengths were adopted from van der Swaelmen et al. (2013).

The barium abundance was measured using the \ion{Ba}{II}~6141.7~{\AA} 
and \ion{Ba}{II}~6496.9~{\AA} lines. As well known, the HFS 
(nuclear spin I~$=3/2$) and the isotopic splitting are important effects to 
be taken into account in Ba transitions. Following Barbuy et al. (2014), 
the hyperfine coupling constants were adopted 
experimentally from Rutten (1978) and Biehl (1976). 
According to Asplund et al. (2009), the major contribution comes from the 
isotope $^{138}$Ba (71.698\%), followed 
by $^{137}$Ba (11.232\%), $^{136}$Ba (7.854\%), $^{135}$Ba (6.592\%), and 
$^{134}$Ba (2.417\%). The isotopes $^{130}$Ba and $^{132}$Ba together represent 
less than 0.11\% and they were ignored in the computations. 
In addition, to compute the profile for \ion{Ba}{II}~6141.7~{\AA}, it is 
important to include a blend with the \ion{Fe}{I}~6141.7~{\AA}, 
for which we adopted \loggf~$=-1.60$ (Barbuy et al. 2014). 
The fit to this line for star 230 is shown in 
Fig. \ref{Y_Ba_La_Eu} (upper right panel).

The lanthanum abundance is a contribution of two stable isotopes. 
The most relevant is $^{139}$La, with $99.909\%$ in the solar material, 
and the only isotope included in the computations since $^{138}$La 
contributes less than $0.1\%$ (Asplund et al. 2009). 
The HFS values were computed with coupling constants A and B adopted from 
Lawler et al. (2001a) and Biehl (1976), with nuclear spin I~$=7/2$. 
The final abundances are based on three \ion{La}{II} lines, located 
at 6320.4~{\AA}, 6390.5~{\AA}, and 6774.3~{\AA}. 
In Fig. \ref{Y_Ba_La_Eu} we show the fit to the \ion{La}{II}~6320.5~{\AA} 
line for star 224 (lower left panel). 

Europium is the heaviest element measured in the sample stars. 
The solar isotopic fraction $^{151}$Eu~$=47.81\%$ and $^{153}$Eu~$=52.19\%$ 
(Asplund et al. 2009) was adopted, with nuclear spin I~$=5/2$. 
We computed the HFS using coupling constants A and B from Lawler et al. (2001b). 
The final europium abundances were derived from the \ion{Eu}{II}~6437.6~{\AA} 
and \ion{Eu}{II}~6645.1~{\AA} lines, and in Fig. \ref{Y_Ba_La_Eu} we show 
(lower right panel) the result for the \ion{Eu}{II}~6437.6~{\AA} line 
in star 238.

\subsection{Uncertainties on the derived abundances}

The typical errors in the spectroscopic atmospheric parameters are 
$\Delta$\Teff~$=100$~K, $\Delta$\logg~$=0.1$~dex, and $\Delta\xi=0.2$~\kms. 
Since the stellar parameters are not independent, the quadratic sum of 
the abundance uncertainties that arise from each of these three sources 
independently will add significant covariance terms to the final error budget.

We solved this problem by creating a new atmospheric model with a $100$~K lower 
temperature, determining the corresponding surface gravity \logg~and microturbulent 
velocity $\xi$ by the spectroscopic method. The difference between the 
abundances derived with this new model and the nominal model in each 
star are expected to represent the total error budget arising from the 
stellar parameters. 

The observational uncertainties are assumed as the standard error of the mean 
obtained with the abundances from individual lines. For elements with three or 
less lines used to determine the average, we adopted the Fe observational 
error as a representative value. The final error is the quadratic sum of the 
uncertainty from the atmospheric parameters and the observational error. 
Table \ref{error} shows the results in star 235 as an example. 
It is important to note, as described already in Table \ref{param}, that 
the observation errors in stars 221, 224, and 230 are significantly larger
in comparison with stars 235 and 238, as a consequence of differences 
in the S/N.

\begin{table}
\caption{Observational and atmospheric errors in star 235, as well as the final uncertainties.}             
\label{error}      
\centering                          
\begin{tabular}{ccccc}        
\hline\hline                 
\noalign{\smallskip}
\hbox{Element} & \hbox{$\Delta_{obs}$} & \hbox{$\Delta_{atm}$} & \hbox{$\Delta_{final[X/H]}$} & \hbox{$\Delta_{final[X/Fe]}$} \\
\noalign{\smallskip}
\hline
\noalign{\smallskip}
 & \hbox{(dex)} & \hbox{(dex)} & \hbox{(dex)} & \hbox{(dex)} \\
\noalign{\smallskip} 
\hline
\noalign{\smallskip}
\hbox{\ion{Fe}{I}}  & 0.15 & 0.08 & 0.17 & ---- \\
\hbox{\ion{Fe}{II}} & 0.09 & 0.06 & 0.11 & ---- \\
\hbox{C(C$_{2}$)}   & 0.09 & 0.04 & 0.10 & 0.14 \\
\hbox{N(CN)}        & 0.09 & 0.14 & 0.17 & 0.19 \\
\hbox{[\ion{O}{I}]} & 0.09 & 0.12 & 0.15 & 0.18 \\
\hbox{\ion{Na}{I}}  & 0.09 & 0.02 & 0.09 & 0.14 \\
\hbox{\ion{Mg}{I}}  & 0.09 & 0.03 & 0.09 & 0.14 \\
\hbox{\ion{Al}{I}}  & 0.09 & 0.05 & 0.10 & 0.14 \\
\hbox{\ion{Si}{I}}  & 0.03 & 0.02 & 0.04 & 0.11 \\
\hbox{\ion{Ca}{I}}  & 0.10 & 0.02 & 0.10 & 0.14 \\
\hbox{\ion{Ti}{I}}  & 0.06 & 0.08 & 0.10 & 0.14 \\
\hbox{\ion{Ti}{II}} & 0.04 & 0.07 & 0.08 & 0.13 \\
\hbox{\ion{Y}{I}}   & 0.09 & 0.07 & 0.11 & 0.15 \\
\hbox{\ion{Zr}{I}}  & 0.09 & 0.08 & 0.12 & 0.16 \\
\hbox{\ion{Ba}{II}} & 0.09 & 0.02 & 0.09 & 0.14 \\
\hbox{\ion{La}{II}} & 0.09 & 0.07 & 0.11 & 0.15 \\
\hbox{\ion{Eu}{II}} & 0.09 & 0.05 & 0.10 & 0.14 \\
\noalign{\smallskip}
\hline
\end{tabular}
\end{table}

\section{Discussion}

\begin{figure}
\centering
\resizebox{90mm}{!}{\includegraphics[angle=0]{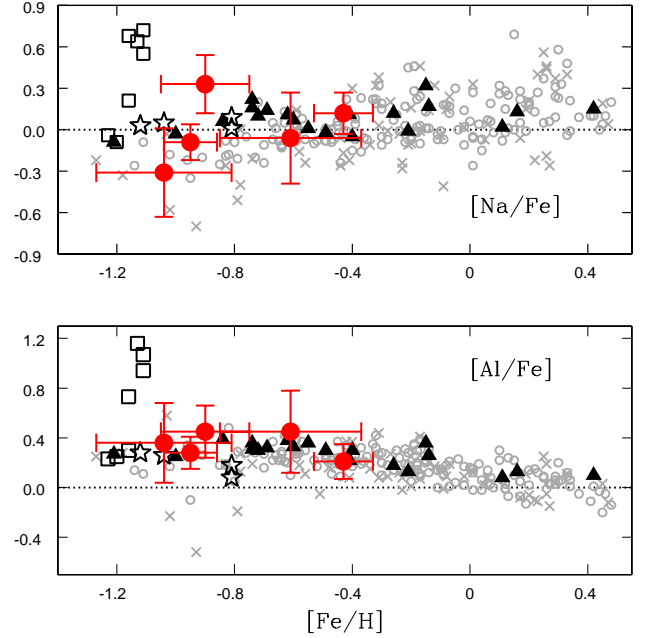}}
\caption{[Na/Fe] (\textbf{upper panel}) and [Al/Fe] (\textbf{lower panel}) 
abundance ratios as a function of the metallicity for the five sample 
stars (filled red circles), compared with literature abundances 
from Yong et al. (2014; open black squares), Barbuy et al. (2014; open 
black stars), Bensby et al. (2013; filled black triangles), 
Johnson et al. (2012; grey crosses), and Johnson et al. (2014; open grey circles).}
\label{Na_Al_compara}
\end{figure}

\begin{figure}
\centering
\resizebox{90mm}{!}{\includegraphics[angle=0]{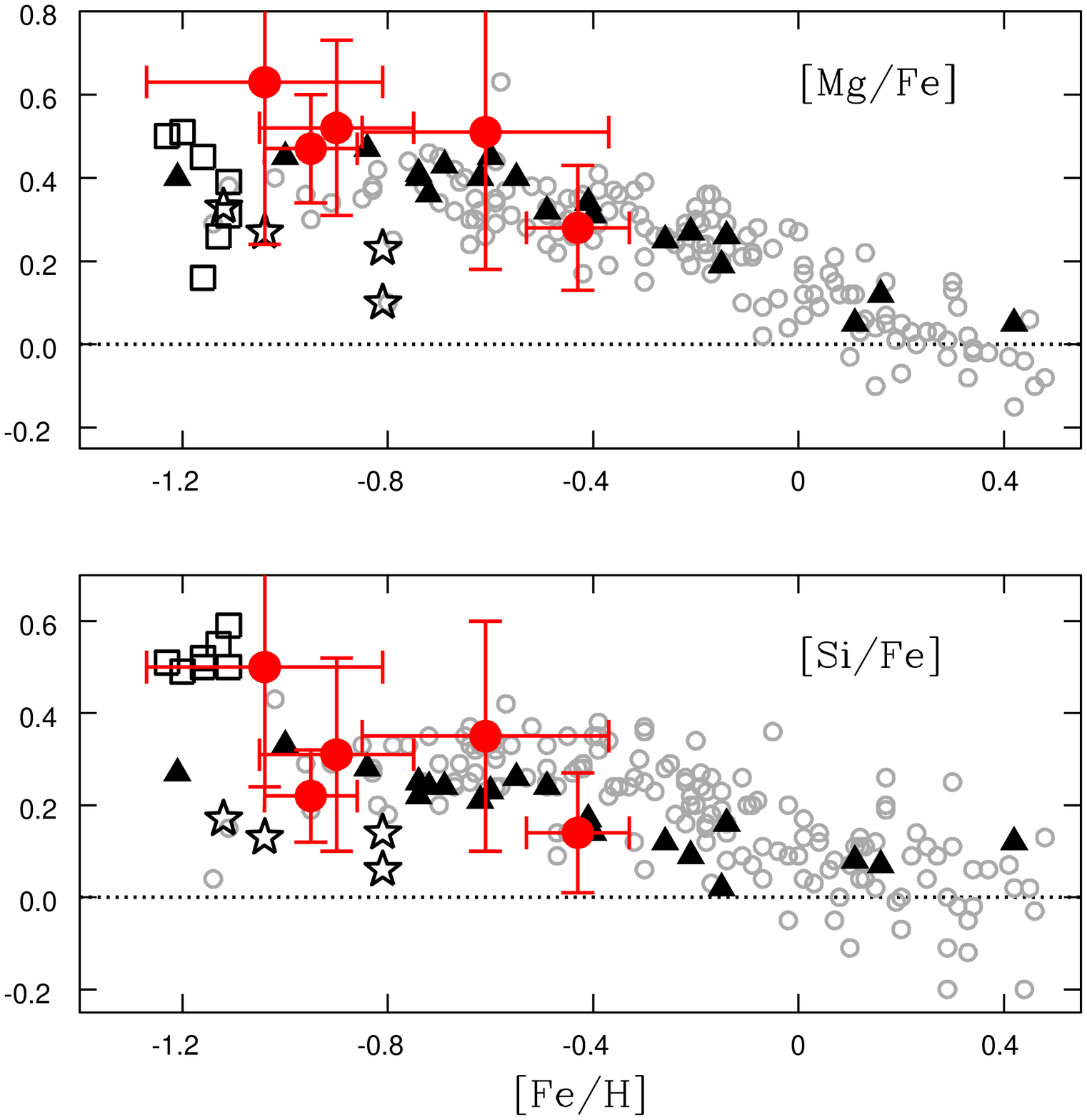}}
\caption{[Mg/Fe] (\textbf{upper panel}) and [Si/Fe] (\textbf{lower panel}) 
abundance ratios as a function of the metallicity for the five sample 
stars, compared with literature abundances. Symbols are the same as 
in Fig. \ref{Na_Al_compara}.}
\label{Mg_Si_compara}
\end{figure}

\begin{figure}
\centering
\resizebox{90mm}{!}{\includegraphics[angle=0]{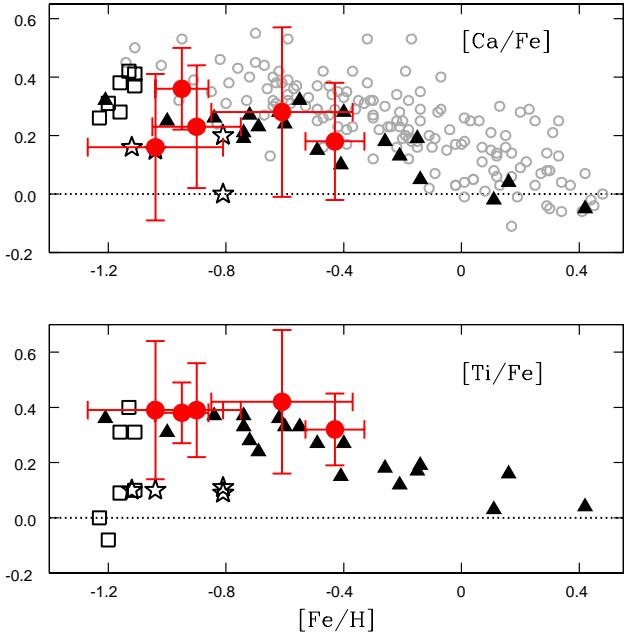}}
\caption{[Ca/Fe] (\textbf{upper panel}) and [Ti/Fe] (\textbf{lower panel}) 
abundance ratios as a function of the metallicity for the five sample 
stars, compared with literature abundances. Symbols are the same as 
in Fig. \ref{Na_Al_compara}.}
\label{Ca_Ti_compara}
\end{figure}

The sample stars analysed at high spectral resolution are confirmed to be 
moderately metal-poor with $-1.04\leq$~[Fe/H]~$\leq-0.43$ 
with no enhancement in [C/Fe], and some stars show high 
nitrogen abundances [N/Fe]: $+0.82\pm0.26$, $+0.71\pm0.34$, $+0.97\pm0.19$, 
and $+0.35\pm0.20$, for 221, 230, 235, and 238, respectively.

In the context of APOGEE (Majewski et al. 2015), Schiavon et al. (2015) recently reported the 
discovery of a population of Galactic bulge field stars 
with high values of [N/Fe], which is correlated with [Al/Fe] 
and anticorrelated with [C/Fe], typical of globular cluster stars 
(Hesser et al. 1982; Gratton, Carretta \& Bragaglia 2012). 
The N-rich stars in our sample could be related to the population 
newly discovered by Schiavon et al. (2015). According to the authors, 
abundance ratios [N/Fe]~$>+0.6$ cannot be 
explained by the CN-cycle mixing scenario, and the contamination by 
mass transfer binaries mechanism can only account for, at most, 
$25$\% of their sample. Possible scenarios for the origin of the 
N-rich stars are: i) dissolution of an early population of globular 
clusters (Belokurov et al. 2006; Shapiro et al. 2010; Kruijssen 2015; Bournaud 2016); 
ii) a shared (or similar) molecular cloud responsible for forming 
these stars and the globular cluster (Longmore et al. 2014; Schiavon et al. 2015); 
iii) these stars are among the oldest in the Galaxy and their abundances are imprints 
of the first stellar generations (Tumlinson 2010; Chiappini et al. 2011). 

To better place our results in the context of the Galactic 
bulge, we present comparisons with literature abundances in 
bulge stars. As already described in Sect. 5.1, we adopted results from 
Yong et al. (2014) of seven stars in the globular cluster M~62, 
the ninth most luminous Galactic globular cluster, which also 
presents an extended horizontal branch. The stars 
were observed with the High Dispersion Spectrograph
(HDS; Noguchi et al. 2002) on the Subaru Telescope 
and with the Magellan Inamori Kyocera Echelle spectrograph (MIKE; Bernstein et al. 2003) at 
the Magellan-II Telescope. The authors found a 
good agreement between the scaled-solar r-process distribution 
and the derived abundances for the elements heavier than La, 
as well as an enhancement in Y, Zr, and Ba in comparison 
with the solar r-process pattern. According to Yong et al. (2014), 
these results are incompatible with the s-process in AGB stars 
and suggest the fast-rotating massive stars as a possible 
solution.

Also discussed in Sect. 5.1, Barbuy et al. (2014) analysed 
four stars in the globular cluster NGC~6522, which 
appears to be the oldest known Milky Way globular cluster. 
The targets were observed at the VLT using the UVES spectrograph (Dekker et al. 2000) 
in FLAMES-UVES mode. They found an enhancement in 
s-process-dominant elements, suggesting spinstars as a possibility to form these elements, 
besides the usual explanations of mass transfer from s-process-rich AGB stars and extra 
mechanisms as the weak r-process as possible scenarios to explain the abundance signatures. 
Ness et al. (2014) found similar results to Barbuy et al. (2014), but these authors insist 
that the abundances of this cluster were measured to be similar to bulge field stars, halo stars, 
and other Galactic globular clusters of the same metallicity. We note that NGC~6522 appears to be 
among the oldest globular clusters, and as such it should show signatures as one of the main pieces 
of the sub-systems that first formed in the central parts of the Galaxy.

In addition, we selected 62 red giant stars analysed in 
Johnson et al. (2012), observed in Plaut's low-extinction 
window. Using the Hydra multi-fiber spectrograph on the CTIO Blanco 
$4$~m telescope, the stars were observed at 
$l=-1^{\circ}$ and $b=-8.5^{\circ}$ (field 1) and at 
$l=0^{\circ}$ and $b=-8^{\circ}$ (field 2). 
Another 156 red giant branch stars in two Galactic bulge fields 
centred near $l=+5.25^{\circ}$ and $b=-3.02^{\circ}$ and 
$l=0^{\circ}$ and $b=-12^{\circ}$ analysed in 
Johnson et al. (2014), using FLAMES-GIRAFFE spectra, were 
selected in the comparison.

In Bensby et al. (2013), 58 microlensed bulge dwarfs and 
subgiants stars were analysed. The authors 
estimated the stellar ages based on isochrones 
(Demarque et al. 2004) and probability distribution 
functions (Bensby et al. 2011), so we selected 
22 stars with ages older than 11 Gyr, avoiding the younger 
stellar populations present in the bulge. 

Finally, 56 other bulge giant stars were selected from 
Van der Swaelmen et al. (2016), who analysed the 
heavy elements in this sample. Already studied 
in Zoccali et al. (2006), Lecureur et al. (2007), 
and Barbuy et al. (2013, 2015), the observations were 
performed with the multi-fibre spectrograph FLAMES-UVES, 
at the UT2 Kuyen VLT/ESO telescope. The stars are located 
at the Baade's Window ($l=1.14^{\circ}$, $b=-4.2^{\circ}$), 
at the Blanco field ($l=0^{\circ}$, $b=-12^{\circ}$), 
at the field of NGC~6553 ($l=5.2^{\circ}$, $b=-3^{\circ}$), 
and in an additional field at $l=0.2^{\circ}$ and $b=-6^{\circ}$.

In Fig. \ref{Na_Al_compara} we show the comparisons for 
the odd-Z elements sodium (upper panel) and aluminum 
(lower panel), while the $\alpha$-elements Mg, Si, Ca, and Ti 
are presented in Figs. \ref{Mg_Si_compara} and \ref{Ca_Ti_compara}. 
As a general behaviour, the derived abundances in the sample 
stars are in agreement with the literature, for objects with 
the same metallicities, within the error bars. 
Chemical similarities between globular cluster primary 
stellar populations and field stars for a given metallicity 
were studied in other works (see e.g. Renzini 2008; 
Gratton et al. 2012; Schiavon et al. 2015).

Chemical inhomogeneity among stars within individual 
globular clusters are well known for elements like C, N, 
O, Na, Mg, and Al (see Kraft 1994; Gratton et al. 2004, 2012; 
M\'esz\'aros et al. 2015), and several models are claimed 
to explain this observational signature (Fenner et al. 2004; 
Decressin et al. 2007; Renzini 2008; Marcolini et al. 2009; 
Hopkins 2014; Renzini et al. 2015). To test these so-called abundance 
anomalies, we checked if the Na~$-$~O, Al~$-$~O, and Al~$-$~Mg 
anti-correlations were present in our sample, as well as 
the Na~$-$~Al correlation, but no significant trend was found. 
The $\alpha$-elements abundaces are enhanced, as typical of 
chemical enrichment from core-collapse supernovae 
(Woosley \& Weaver 1995; Nomoto et al. 2013, and references therein). 

\begin{figure}
\centering
\resizebox{90mm}{!}{\includegraphics[angle=0]{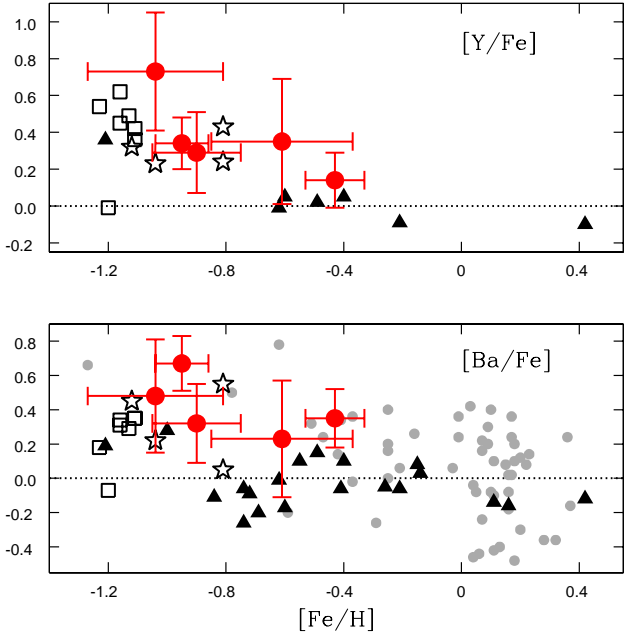}}
\caption{[Y/Fe] (\textbf{upper panel}) and [Ba/Fe] (\textbf{lower panel}) 
abundance ratios as a function of the metallicity for the five sample 
stars, compared with literature abundances. Symbols are the same as 
in Fig. \ref{Na_Al_compara}, in addition to the abundances from 
Van der Swaelmen et al. (2016; filled grey circles).}
\label{Y_Ba_compara}
\end{figure}

\begin{figure}
\centering
\resizebox{90mm}{!}{\includegraphics[angle=0]{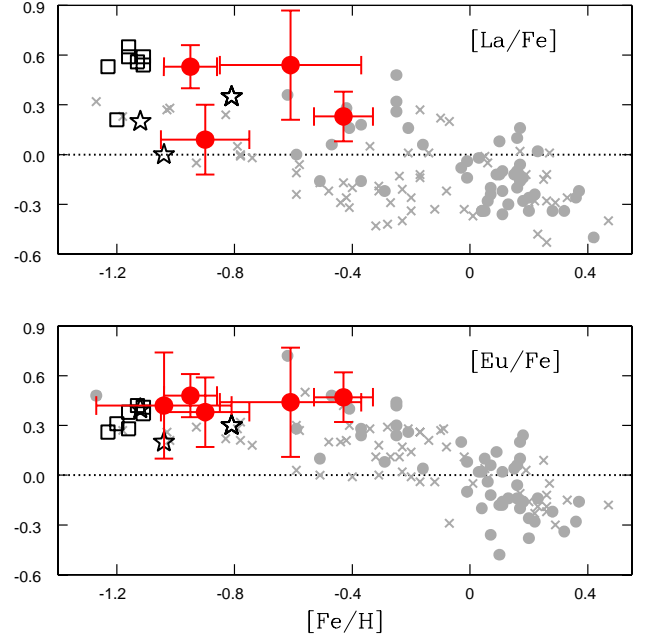}}
\caption{[La/Fe] (\textbf{upper panel}) and [Eu/Fe] (\textbf{lower panel}) 
abundance ratios as a function of the metallicity for the five sample 
stars, compared with literature abundances. Symbols are the same as 
in Figs. \ref{Na_Al_compara} and \ref{Y_Ba_compara}.}
\label{La_Eu_compara}
\end{figure}

Regarding the heavy elements, we derived abundances of Y, Zr, Ba, 
La, and Eu in our sample, and, in Figs. \ref{Y_Ba_compara} and 
\ref{La_Eu_compara}, we show the results in comparison to values 
from literature. A good agreement is observed 
among the selected stars and, analogously to $\alpha$-elements, 
enhancement in the heavy elements was obtained in the sample. 
The only exception to this average behaviour in literature is 
observed in Bensby et al. (2013), for which the major fraction 
of stars shows solar values of [Y/Fe] and [Ba/Fe] abundance ratio.

The behaviour of the [Eu/Fe] abundance ratio is similar with that 
observed for the $\alpha$-elements, which would be expected 
from the main r-process. In the solar material, the r-process is responsible 
for $94\pm0.4$\% of the total Eu abundance (Bisterzo et al. 2014). 
However, most of the Y, Zr, Ba, and La available today in the solar system and in the 
Galaxy has been produced by the s-process in low-mass AGB stars (Sneden et al. 2008, and 
references therein) which, owing to the typical long lifetimes of low-mass stars, would 
not have had time to evolve and pollute the gas before forming the sample stars, which are 
probably very old. Possible scenarios for the enrichment of Y, Zr, Ba, and La derived 
in our sample are: 

i) Early enrichment from r-process only, where an extra mechanism
is required to produce excesses of the lightest
trans-Fe elements with respect to second peak elements
such as Ba and La (e.g. Travaglio et al. 2004; Wanajo et al. 2011; Arcones \& Montes 2011; 
Arcones \& Thielemann 2013; Fujibayashi et al. 2015; Niu et al. 2015);

ii)  s-process elements from  AGB stars bounded in a binary system,
 polluting the observed stars via AGB-mass transfer 
(Bisterzo et al. 2014);

iii) s-process activation in early generations of Spinstars (Meynet et al. 2006; Pignatari et al. 2008; 
Frischknecht et al. 2012; Frischknecht et al. 2015), which pollutes the primordial material before 
forming the oldest bulge field stars. 

\begin{figure}
\centering
\resizebox{90mm}{!}{\includegraphics[angle=0]{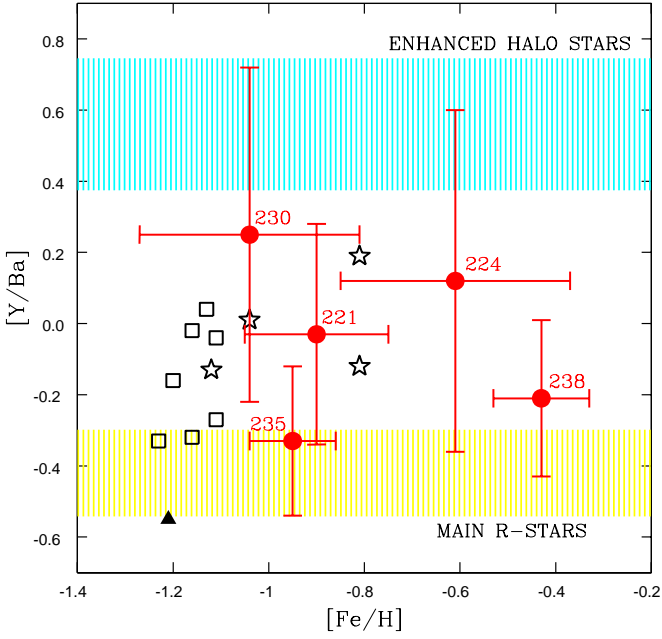}}
\caption{[Y/Ba] vs. [Fe/H] diagram for the sample stars 
and results from literature. Symbols are the same as 
in Figs. \ref{Na_Al_compara}. The yellow and cyan 
regions correspond to the main r-process signature 
and the abundance ratio from metal-poor stars enhanced in 
the lightest heavy elements (see text for details).}
\label{YxBa_compara}
\end{figure}

Figure \ref{YxBa_compara} shows the [Y/Ba] vs. [Fe/H] diagram for bulge 
stars from our sample compared with selected results from the literature. 
From Bensby et al. (2013), we only retained the star enhanced in [Y/Fe]. 
We included the average [Y/Ba]$_{r}=-0.42\pm0.12$ 
abundance ratio value obtained (yellow region) from six halo metal-poor r-element-rich 
stars (HD~221170, HD~115444, CB~22892-052, HE~1523-0901, 
BD~17~3248, and CS~31082-001), compiled in Sneden et al. (2008), 
as a representative value of the main r-process. 
We also show the mean value 
of the [Y/Ba] ratio obtained from six halo metal-poor r-process stars showing enhancement 
in the lightest heavy elements: HD~88609 (Honda et al. 2007), BD~4~2621 (Johnson 2002), 
HD~4306 (Honda et al. 2004), HD~237846 (Roederer et al. 2010), 
HD~122563 (Honda et al. 2007), and HD~140283 (Siqueira-Mello et al. 2015). 
The average value is [Y/Ba]$_{E}=+0.56\pm0.18$ (illustrated in the figure by the cyan 
region - where E stands for ``enhanced''). The source of this enhancement can be manifold 
(see discussion below). We note that the cyan region is meant to only show a reference value since, 
even for the same nucleosynthetic sources, the expected level of enhancement in the bulge and 
halo will be different (see Barbuy et al. 2014).

The figure shows that the value derived in star 235 agrees with those observed in the 
r-element-rich stars, and it can be explained by a pattern arising in a 
so-called main r-process. Considering the error bars, the same conclusion 
could be claimed to explain the [Y/Ba] abundance in star 238. 
On the other hand, the derived abundances in 221, 224, and 230 are barely explained using 
only the main r-process. If the bulge stars with [Fe/H]~$\sim-1$ trace the same early phases of chemical 
enrichment as halo stars with [Fe/H]~$\sim-3$, the sample stars 235 and 
238 may be classified as r-process enhanced stars, analogous to the r-I and r-II metal-poor 
halo stars (Beers \& Christlieb 2005). In fact, using the latter authors' definitions that are 
based on the [Ba/Eu] abundance ratio, objects 235 and 238 must be classified as r/s and r-I stars, 
respectively. This is the first time that these kind of stars are identified in the Galactic bulge.

\begin{figure}
\centering
\resizebox{90mm}{!}{\includegraphics[angle=0]{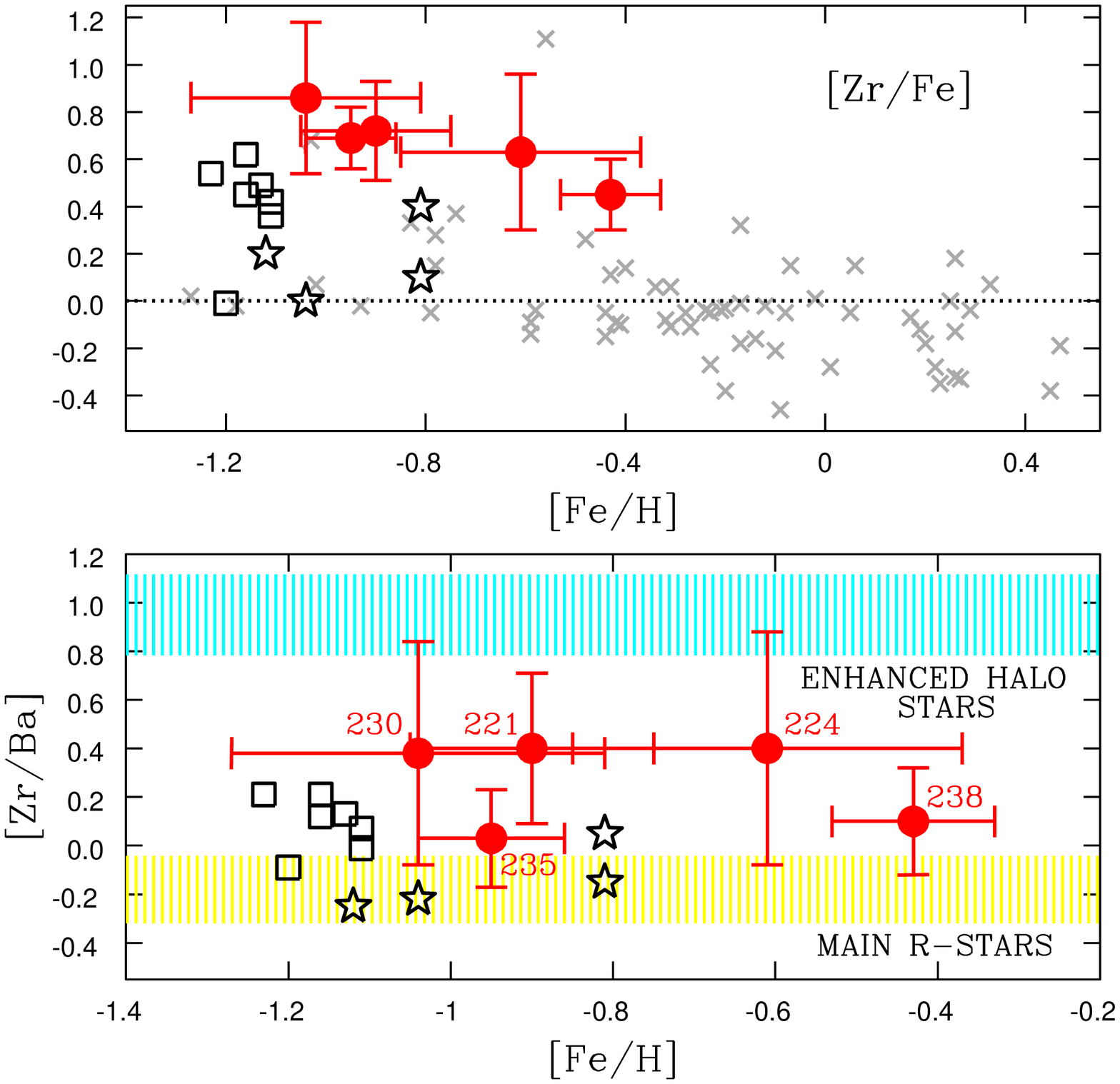}}
\caption{\textbf{Upper panel}: [Zr/Fe] abundance ratios as a 
function of the metallicity for the five sample stars, compared with 
literature abundances. Symbols are the same as in Figs. \ref{Na_Al_compara}. 
\textbf{Lower panel:} [Zr/Ba] vs. [Fe/H] diagram for the 
sample stars and results from literature. Symbols are the same as 
in Figs. \ref{Na_Al_compara}. The yellow and cyan 
regions correspond to the main r-process signature 
and the abundance ratio from metal-poor stars enhanced in 
the lightest heavy elements (see text for details).}
\label{ZrxBa_compara}
\end{figure}

Figure \ref{ZrxBa_compara} presents the same comparison for zirconium. 
In the upper panel the [Zr/Fe] abundance ratios derived in the sample 
are compared with results from the literature. Johnson et al. (2012) 
identified evidence of two separate sequences: a group of stars enhanced 
in [Zr/Fe], and another group moderately poor. Clearly our sample stars 
are members of the enhanced group. In the lower panel we show the 
[Zr/Ba] vs. [Fe/H] diagram for the sample stars and the selected results from 
literature. The selected metal-poor r-element-rich and the enhanced stars 
in lightest heavy elements were also used to define [Zr/Ba]$_{r}=-0.18\pm0.12$ 
(yellow region) and [Zr/Ba]$_{E}=+0.95\pm0.15$ (cyan region), respectively. 
The figure shows that the Zr abundances derived in stars 235 and 238 also agree 
with those observed in the r-element-rich stars, while stars 221, 224, and 230 
require extra mechanism(s) to explain the abundance ratios.

For the Galactic halo, Roederer et al. (2010) show several metal-poor stars located 
in the region between these two extremes abundance regimes, suggesting also a continuous 
range of r-process nucleosynthesis patterns. On the other hand, 
Niu et al. (2015) more recently suggest that the weak r-process 
and the main r-process are two distinct astrophysical processes.

The fundamental challenge that we are facing is that in the early galaxy a 
multitude of different nucleosynthesis processes may have contributed to the production 
of the elements at the first neutron-magic peak beyond iron, including Sr, Y, and Zr. 
Together with the s-process in fast rotating massive stars (Frischknecht et al. 2016) and the weak r-process 
(e.g., Farouqi et al. 2009), other sources could be at play such as: the electron capture supernovae 
(e.g., Wanajo et al. 2011), or the $\alpha$-rich freezout in most energetic core-collapse 
supernovae (e.g., Woosley \& Hoffman 1992), and the intermediary neutron capture i-process 
(Dardelet et al. 2015, and references therein). In addition, neutrino-winds in core-collapse supernovae can 
host a large variety of processes that can produce elements in the same mass region 
(e.g., Fr\"ohlich et al. 2006, Farouqi et al. 2010, Roberts et al. 2010, Arcones \& Montes 2011). 
It is thus crucial to measure as many heavy elements as possible to isolate the different nucleosynthesis sources.

\section{Conclusions}

We have carried out a pilot project with the goal of providing detailed abundances 
for moderately metal-poor Galactic bulge stars that are believed to host imprints left by the first 
stellar generations. In this work, we were able to obtain detailed abundances for five moderately 
metal-poor and [$\alpha$/Fe]~$>0$ stars from one field of the ARGOS survey. Our high-resolution 
FLAMES-UVES spectra have confirmed three out of five stars to have metallicities [Fe/H]~$<-0.8$. 
All stars are confirmed to be $\alpha$-enhanced: overabundances of the typical 
$\alpha$'s Mg, Si, Ca, Ti, and of the odd-Z element Al are clearly detected. 

Three sample stars  exhibit high [N/Fe] abundance ratios. Similar high-[N/Fe] bulge stars 
were recently found in APOGEE (see Schiavon et al. 2015). According to the latter authors, abundance 
ratios [N/Fe]~$>+0.6$ cannot be explained by the CN-cycle mixing scenario. 

The sample stars show enhancements in [Y/Fe], [Zr/Fe], [Ba/Fe], [La/Fe], and [Eu/Fe]. We found that 
three of our stars also show [Y/Ba] and [Zr/Ba] ratios slightly higher than expected from a pure main 
r-process nucleosynthesis. These results are very similar to the recent reported chemical pattern found in some stars of the 
oldest Milky Way globular cluster, NGC~6522 (Barbuy et al. 2014). 
Considering the sample stars as an old population, whereas an enhancement in Eu would be expected 
from the rapid neutron capture process (or r-process), consistent with the observed 
[$\alpha$/Fe] enhancements, the observed anomalous enrichment of the 
dominantly s-process elements [Y/Ba] and [Zr/Ba] in these stars are more difficult to understand using 
standard nucleosynthesis processes.

Finally, previous to the present work, some excesses of s-process-typical elements in the Galactic bulge 
had been found only in globular clusters (Barbuy et al. 2009; Chiappini et al. 2011; Barbuy et al. 2014; Yong et al. 2014). 
The goal of our pilot project was to also look for the existence of these stars in the field. 
Although our sample is very small, three of our stars seem to show not only excesses of the lightest heavy elements with respect to iron, 
but also enhancement in the [Y/Ba] and [Zr/Ba] abundance ratios. The s-process activation in fast-rotating massive stars 
and/or other extra mechanisms are possible solutions to these anomalies. There is a debate in the literature
about the origin of heavy elements in the oldest stars, such that
future large samples are urgently needed to futher explore the impact of these findings in our understanding of 
the nature of the first stellar generations. 


\begin{acknowledgements}
CS, BB, and EC acknowledge grants from CAPES, CNPq and FAPESP. 
MP acknowledges support from NSF grants PHY 02-16783 and PHY 09-22648 
(Joint Institute for Nuclear Astrophysics, JINA), NSF grant PHY-1430152 
(JINA Center for the Evolution of the Elements) and EU MIRG-CT-2006-046520. 
The continued work on codes and in disseminating data is made possible through 
funding from STFC and EU-FP7-ERC-2012-St Grant 306901 (RH, UK). 
MP acknowledges support from the Lendulet-2014 Programme of the Hungarian Academy of 
Sciences and from SNF (Switzerland). 
The research leading to these results has received funding from 
the European Research Council under the European Union's Seventh 
Framework Programme (FP/2007-2013) / ERC Grant Agreement n. 306901. 
R. Hirschi acknowledges support from the World Premier International 
Research Center Initiative (WPI Initiative), MEXT, Japan. 
This work has made use of the VALD database, operated at 
Uppsala University, the Institute of Astronomy RAS in Moscow, 
and the University of Vienna.
\end{acknowledgements}


\begin{appendix} 
\section{Line lists}

\scalefont{0.58}
\begin{longtable}{crcccccccccccccc}
\caption{\label{EW_measurements} Equivalent widths (EW) measured and used to derive new atmospheric parameters 
and iron abundances.}\\
\hline\hline
\noalign{\smallskip}
\hbox{} & \hbox{} & \hbox{} & \hbox{} & \hbox{} & \hbox{} &
\multicolumn{2}{c}{Star 221} & \multicolumn{2}{c}{Star 224} & \multicolumn{2}{c}{Star 230} & \multicolumn{2}{c}{Star 235} & \multicolumn{2}{c}{Star 238} \\
\hbox{Species} &  \hbox{$\lambda$({\AA})} & \hbox{$\chi$$_{ex}$(eV)} & \hbox{log $gf$VALD} & \hbox{log $gf$NIST} & \hbox{log $gf$Adopted} &
\hbox{EW (m{\AA})} & \hbox{A(Fe)} & \hbox{EW (m{\AA})} & \hbox{A(Fe)} & \hbox{EW (m{\AA})} & \hbox{A(Fe)} & \hbox{EW (m{\AA})} & \hbox{A(Fe)} & \hbox{EW (m{\AA})} & \hbox{A(Fe)} \\
\hline
\noalign{\smallskip}
\endfirsthead
\caption{continued.}\\
\hline\hline
\noalign{\smallskip}
\hbox{} & \hbox{} & \hbox{} & \hbox{} & \hbox{} & \hbox{} &
\multicolumn{2}{c}{Star 221} & \multicolumn{2}{c}{Star 224} & \multicolumn{2}{c}{Star 230} & \multicolumn{2}{c}{Star 235} & \multicolumn{2}{c}{Star 238} \\
\hbox{Species} &  \hbox{$\lambda$({\AA})} & \hbox{$\chi$$_{ex}$(eV)} & \hbox{log $gf$VALD} & \hbox{log $gf$NIST} & \hbox{log $gf$Adopted} &
\hbox{EW (m{\AA})} & \hbox{A(Fe)} & \hbox{EW (m{\AA})} & \hbox{A(Fe)} & \hbox{EW (m{\AA})} & \hbox{A(Fe)} & \hbox{EW (m{\AA})} & \hbox{A(Fe)} & \hbox{EW (m{\AA})} & \hbox{A(Fe)} \\
\hline
\noalign{\smallskip}
\endhead
\hline
\endfoot
\hbox{Fe I}  & 6137.691 & 2.588 & $-$1.403 & $-$1.403 & $-$1.403 & ---- & ---- &113.5 & 6.62 & 73.4 & 5.80 &159.3 & 6.86 &156.9 & 6.97 \\
\hbox{Fe I}  & 6151.618 & 2.176 & $-$3.299 & $-$3.299 & $-$3.299 & 52.4 & 6.17 & 56.2 & 6.88 & ---- & ---- & 61.7 & 6.42 & 89.8 & 7.27 \\
\hbox{Fe I}  & 6157.728 & 4.076 & $-$1.260 & $-$1.220 & $-$1.220 & 70.8 & 6.93 & 48.4 & 6.77 & 52.5 & 6.80 & 57.6 & 6.62 & 70.2 & 7.08 \\
\hbox{Fe I}  & 6159.377 & 4.608 & $-$1.970 & -------- & $-$1.970 & ---- & ---- & ---- & ---- &  8.0 & 6.85 & ---- & ---- & ---- & ---- \\
\hbox{Fe I}  & 6165.360 & 4.143 & $-$1.474 & $-$1.474 & $-$1.474 & 83.1 & 7.56 & 44.6 & 7.01 & 20.8 & 6.34 & 30.4 & 6.36 & 50.3 & 6.95 \\
\hbox{Fe I}  & 6173.335 & 2.223 & $-$2.880 & $-$2.880 & $-$2.880 & 78.2 & 6.39 & 46.2 & 6.27 & 32.8 & 5.86 & 90.3 & 6.69 & ---- & ---- \\
\hbox{Fe I}  & 6180.204 & 2.727 & $-$2.586 & $-$2.649 & $-$2.649 & 73.3 & 6.70 & 36.6 & 6.40 & 19.9 & 5.87 & ---- & ---- & 79.3 & 7.07 \\
\hbox{Fe I}  & 6187.990 & 3.943 & $-$1.720 & $-$1.670 & $-$1.670 & 49.9 & 6.71 & 22.0 & 6.42 & ---- & ---- & 42.7 & 6.58 & 57.2 & 7.05 \\
\hbox{Fe I}  & 6200.313 & 2.608 & $-$2.437 & $-$2.437 & $-$2.437 & 47.6 & 5.75 & ---- & ---- & 64.3 & 6.63 & 69.8 & 6.28 & ---- & ---- \\
\hbox{Fe I}  & 6213.430 & 2.223 & $-$2.482 & $-$2.482 & $-$2.482 &102.9 & 6.54 & 94.6 & 6.97 & 82.0 & 6.64 &105.8 & 6.61 &121.8 & 7.11 \\
\hbox{Fe I}  & 6219.281 & 2.198 & $-$2.433 & $-$2.433 & $-$2.433 & ---- & ---- & 92.5 & 6.85 & 88.9 & 6.71 &114.3 & 6.70 &141.8 & 7.32 \\
\hbox{Fe I}  & 6220.783 & 3.882 & $-$2.460 & -------- & $-$2.460 &  9.8 & 6.33 & ---- & ---- & ---- & ---- & 11.8 & 6.47 & 26.2 & 7.07 \\
\hbox{Fe I}  & 6226.736 & 3.882 & $-$2.220 & -------- & $-$2.220 & 49.3 & 7.17 & ---- & ---- & ---- & ---- & 15.3 & 6.37 & ---- & ---- \\
\hbox{Fe I}  & 6229.228 & 2.845 & $-$2.805 & $-$2.805 & $-$2.805 & 56.0 & 6.60 & 64.5 & 7.35 & 42.1 & 6.72 & 50.7 & 6.53 & 50.8 & 6.72 \\
\hbox{Fe I}  & 6240.646 & 2.223 & $-$3.233 & $-$3.173 & $-$3.173 & 36.7 & 5.77 & 52.0 & 6.71 & 41.5 & 6.36 & 58.0 & 6.28 & 80.2 & 7.01 \\
\hbox{Fe I}  & 6246.318 & 3.602 & $-$0.733 & $-$0.877 & $-$0.877 &102.5 & 6.65 & 98.2 & 6.80 & 66.4 & 6.18 & 92.5 & 6.43 & ---- & ---- \\
\hbox{Fe I}  & 6252.555 & 2.404 & $-$1.687 & $-$1.687 & $-$1.687 &133.1 & 6.59 &113.5 & 6.70 & 78.8 & 5.99 &133.4 & 6.58 &150.3 & 6.97 \\
\hbox{Fe I}  & 6254.258 & 2.279 & $-$2.443 & $-$2.426 & $-$2.426 &114.0 & 6.92 & 68.7 & 6.43 &103.7 & 7.13 &146.7 & 7.49 &129.3 & 7.38 \\
\hbox{Fe I}  & 6265.133 & 2.176 & $-$2.550 & $-$2.550 & $-$2.550 &101.8 & 6.52 & 76.5 & 6.60 & 49.0 & 5.85 &104.7 & 6.59 &115.6 & 7.02 \\
\hbox{Fe I}  & 6270.225 & 2.858 & $-$2.464 & $-$2.609 & $-$2.609 & 46.6 & 6.21 & 40.4 & 6.59 & ---- & ---- & 61.0 & 6.56 & 72.7 & 7.03 \\
\hbox{Fe I}  & 6271.278 & 3.332 & $-$2.703 & $-$2.703 & $-$2.703 & 25.8 & 6.43 & ---- & ---- & 32.9 & 6.95 & 21.3 & 6.37 & 49.1 & 7.16 \\
\hbox{Fe I}  & 6297.793 & 2.223 & $-$2.740 & $-$2.740 & $-$2.740 &100.5 & 6.74 & 68.0 & 6.64 & 71.8 & 6.65 & ---- & ---- &122.7 & 7.38 \\
\hbox{Fe I}  & 6301.501 & 3.654 & $-$0.718 & $-$0.718 & $-$0.718 & 77.5 & 6.02 & ---- & ---- & ---- & ---- & 97.1 & 6.41 &103.8 & 6.68 \\
\hbox{Fe I}  & 6302.493 & 3.686 & $-$0.973 & -------- & $-$0.973 & ---- & ---- & ---- & ---- & 61.6 & 6.26 & 87.5 & 6.51 & ---- & ---- \\
\hbox{Fe I}  & 6311.500 & 2.832 & $-$3.141 & $-$3.141 & $-$3.141 & 61.2 & 7.03 & ---- & ---- & ---- & ---- & 35.0 & 6.52 & 53.8 & 7.10 \\
\hbox{Fe I}  & 6315.306 & 4.143 & $-$1.232 & $-$1.232 & $-$1.232 & 60.7 & 6.77 & 70.1 & 7.34 & 39.3 & 6.57 & 65.3 & 6.88 & 77.9 & 7.33 \\
\hbox{Fe I}  & 6315.811 & 4.076 & $-$1.710 & $-$1.660 & $-$1.660 & 43.7 & 6.72 & 45.5 & 7.14 & 37.2 & 6.87 & 37.4 & 6.61 & 39.5 & 6.81 \\
\hbox{Fe I}  & 6322.685 & 2.588 & $-$2.426 & $-$2.426 & $-$2.426 & 86.4 & 6.60 & 84.1 & 7.11 & ---- & ---- & 92.4 & 6.75 &104.1 & 7.20 \\
\hbox{Fe I}  & 6335.330 & 2.198 & $-$2.177 & $-$2.177 & $-$2.177 & 94.6 & 6.01 & 88.6 & 6.50 & 79.6 & 6.24 &119.0 & 6.50 &130.8 & 6.89 \\
\hbox{Fe I}  & 6336.824 & 3.686 & $-$0.856 & $-$0.856 & $-$0.856 & 80.6 & 6.26 & 74.5 & 6.47 & 64.8 & 6.21 & 88.8 & 6.42 & 95.2 & 6.70 \\
\hbox{Fe I}  & 6344.149 & 2.433 & $-$2.923 & $-$2.923 & $-$2.923 & 77.5 & 6.68 & ---- & ---- & 48.7 & 6.51 & 82.2 & 6.82 &104.9 & 7.53 \\
\hbox{Fe I}  & 6355.029 & 2.845 & $-$2.350 & $-$2.291 & $-$2.291 & 83.6 & 6.72 & 78.8 & 7.15 & 75.8 & 7.01 & 86.5 & 6.79 &106.6 & 7.41 \\
\hbox{Fe I}  & 6380.743 & 4.186 & $-$1.376 & $-$1.376 & $-$1.376 & 56.5 & 6.86 & 30.9 & 6.63 & 32.9 & 6.61 & 37.7 & 6.46 & 61.7 & 7.15 \\
\hbox{Fe I}  & 6392.538 & 2.279 & $-$4.030 & -------- & $-$4.030 & 42.9 & 6.81 & 45.0 & 7.44 & ---- & ---- & 30.9 & 6.63 & 39.7 & 7.00 \\
\hbox{Fe I}  & 6393.601 & 2.433 & $-$1.432 & $-$1.576 & $-$1.576 &112.8 & 6.12 & 88.3 & 6.17 & 87.0 & 6.07 &143.0 & 6.61 &161.6 & 6.98 \\
\hbox{Fe I}  & 6408.018 & 3.686 & $-$1.018 & $-$1.018 & $-$1.018 & 69.1 & 6.17 & 77.1 & 6.77 & 67.2 & 6.49 & 83.8 & 6.51 & 93.1 & 6.89 \\
\hbox{Fe I}  & 6411.649 & 3.653 & $-$0.595 & $-$0.718 & $-$0.718 & 95.5 & 6.39 & 96.6 & 6.66 & 69.8 & 6.14 & 98.6 & 6.43 &104.3 & 6.67 \\
\hbox{Fe I}  & 6419.949 & 4.733 & $-$0.240 & $-$0.270 & $-$0.270 & 77.6 & 6.88 & ---- & ---- & 75.1 & 7.00 & 57.8 & 6.44 & ---- & ---- \\
\hbox{Fe I}  & 6421.350 & 2.279 & $-$2.027 & $-$2.027 & $-$2.027 &148.3 & 7.13 &141.5 & 7.28 & 92.7 & 6.50 &130.6 & 6.79 &136.6 & 7.06 \\
\hbox{Fe I}  & 6430.846 & 2.176 & $-$2.006 & $-$2.006 & $-$2.006 &108.4 & 6.08 &121.0 & 6.83 & 77.3 & 5.98 &138.4 & 6.59 &149.3 & 6.92 \\
\hbox{Fe I}  & 6469.193 & 4.835 & $-$0.770 & $-$0.810 & $-$0.810 & 71.2 & 7.40 & ---- & ---- & ---- & ---- & ---- & ---- & 56.7 & 7.21 \\
\hbox{Fe I}  & 6475.624 & 2.559 & $-$2.942 & $-$2.942 & $-$2.942 & 74.4 & 6.78 & 59.7 & 7.02 & 62.4 & 7.01 & 73.2 & 6.78 & 84.4 & 7.25 \\
\hbox{Fe I}  & 6481.870 & 2.279 & $-$2.984 & $-$2.984 & $-$2.984 & 87.5 & 6.74 & 73.5 & 7.05 & 54.1 & 6.51 & 74.6 & 6.48 & 93.4 & 7.11 \\
\hbox{Fe I}  & 6494.980 & 2.404 & $-$1.273 & $-$1.273 & $-$1.273 &149.0 & 6.35 &123.8 & 6.40 &123.4 & 6.33 & ---- & ---- &165.9 & 6.68 \\
\hbox{Fe I}  & 6498.939 & 0.958 & $-$4.699 & $-$4.687 & $-$4.687 &119.9 & 7.55 & 80.1 & 7.40 & ---- & ---- & 86.5 & 6.77 &101.1 & 7.38 \\
\hbox{Fe I}  & 6518.367 & 2.832 & $-$2.460 & $-$2.298 & $-$2.298 & 66.5 & 6.29 & 55.8 & 6.59 & 45.6 & 6.27 & 73.6 & 6.47 & 73.2 & 6.67 \\
\hbox{Fe I}  & 6533.929 & 4.558 & $-$1.460 & $-$1.430 & $-$1.430 & ---- & ---- & ---- & ---- & ---- & ---- & 22.4 & 6.59 & ---- & ---- \\
\hbox{Fe I}  & 6546.239 & 2.759 & $-$1.536 & $-$1.536 & $-$1.536 &114.7 & 6.51 & 85.7 & 6.43 & 79.0 & 6.22 &114.3 & 6.49 &137.9 & 7.04 \\
\hbox{Fe I}  & 6569.215 & 4.733 & $-$0.420 & $-$0.450 & $-$0.450 & 54.7 & 6.54 & 59.9 & 6.93 & 38.1 & 6.40 & 54.2 & 6.54 & 80.7 & 7.23 \\
\hbox{Fe I}  & 6574.228 & 0.990 & $-$5.023 & $-$5.004 & $-$5.004 & 59.0 & 6.44 & 68.1 & 7.45 & 57.8 & 7.10 & ---- & ---- & 81.6 & 7.26 \\
\hbox{Fe I}  & 6575.015 & 4.733 & $-$2.710 & $-$2.710 & $-$2.710 & 83.4 & 6.78 & 67.0 & 6.99 & 58.1 & 6.69 & 76.0 & 6.63 & 92.4 & 7.21 \\
\hbox{Fe I}  & 6581.210 & 1.485 & $-$4.679 & $-$4.679 & $-$4.679 & 53.9 & 6.66 & 43.7 & 7.13 & ---- & ---- & ---- & ---- & ---- & ---- \\
\hbox{Fe I}  & 6593.870 & 2.433 & $-$2.422 & $-$2.422 & $-$2.422 & ---- & ---- & ---- & ---- & 77.8 & 6.70 &103.1 & 6.74 & ---- & ---- \\
\hbox{Fe I}  & 6597.561 & 4.795 & $-$1.070 & $-$1.050 & $-$1.050 & 76.0 & 7.69 & 55.8 & 7.50 & 17.7 & 6.54 & 24.0 & 6.54 & ---- & ---- \\
\hbox{Fe I}  & 6608.026 & 2.279 & $-$4.030 & -------- & $-$4.030 & 62.6 & 7.21 & 46.0 & 7.45 & ---- & ---- & 40.6 & 6.81 & 47.4 & 7.15 \\
\hbox{Fe I}  & 6609.110 & 2.559 & $-$2.692 & $-$2.692 & $-$2.692 & 85.6 & 6.77 & 81.6 & 7.26 & 27.7 & 5.91 & 81.6 & 6.70 &101.6 & 7.35 \\
\hbox{Fe I}  & 6627.544 & 4.548 & $-$1.680 & -------- & $-$1.680 & 24.0 & 6.84 & ---- & ---- & ---- & ---- & ---- & ---- & 27.8 & 7.10 \\
\hbox{Fe I}  & 6677.986 & 2.692 & $-$1.418 & $-$1.418 & $-$1.418 &128.8 & 6.53 & ---- & ---- & 95.2 & 6.35 &139.8 & 6.68 &138.6 & 6.84 \\
\hbox{Fe I}  & 6699.141 & 4.593 & $-$2.101 & $-$2.101 & $-$2.101 & ---- & ---- & ---- & ---- & ---- & ---- & ---- & ---- &  7.3 & 6.83 \\
\hbox{Fe I}  & 6703.567 & 2.759 & $-$3.160 & $-$3.060 & $-$3.060 & 81.6 & 7.29 & 56.7 & 7.28 & ---- & ---- & 46.5 & 6.56 & 68.6 & 7.23 \\
\hbox{Fe I}  & 6705.102 & 4.607 & $-$1.392 & -------- & $-$1.392 & 54.7 & 7.32 & 39.2 & 7.28 & ---- & ---- & 30.9 & 6.82 & 45.4 & 7.27 \\
\hbox{Fe I}  & 6710.319 & 1.485 & $-$4.880 & -------- & $-$4.880 & 54.2 & 6.86 & 45.8 & 7.37 & ---- & ---- & 32.4 & 6.49 & 57.6 & 7.22 \\
\hbox{Fe I}  & 6713.744 & 4.795 & $-$1.600 & -------- & $-$1.600 & ---- & ---- & ---- & ---- & ---- & ---- & 10.3 & 6.61 & 23.2 & 7.19 \\
\hbox{Fe I}  & 6715.383 & 4.608 & $-$1.640 & -------- & $-$1.640 & ---- & ---- & ---- & ---- & 17.0 & 6.89 & 12.0 & 6.51 & 35.0 & 7.30 \\
\hbox{Fe I}  & 6725.357 & 4.103 & $-$2.300 & -------- & $-$2.300 & ---- & ---- & 14.3 & 6.96 & 16.4 & 6.96 & 17.7 & 6.77 & 21.4 & 7.02 \\
\hbox{Fe I}  & 6726.666 & 4.607 & $-$1.133 & -------- & $-$1.133 & ---- & ---- & 52.7 & 7.31 & 31.8 & 6.78 & 31.5 & 6.57 & 46.7 & 7.04 \\
\hbox{Fe I}  & 6733.151 & 4.638 & $-$1.580 & -------- & $-$1.580 & ---- & ---- & 16.0 & 6.90 & ---- & ---- & 17.4 & 6.68 & ---- & ---- \\
\hbox{Fe I}  & 6739.521 & 1.557 & $-$4.794 & $-$4.794 & $-$4.794 & 47.0 & 6.71 & 23.8 & 6.85 & ---- & ---- & 34.8 & 6.55 & 38.0 & 6.82 \\
\hbox{Fe I}  & 6752.707 & 4.638 & $-$1.204 & $-$1.204 & $-$1.204 & 24.7 & 6.48 & ---- & ---- & ---- & ---- & 11.6 & 6.09 & ---- & ---- \\
\hbox{Fe II} & 6084.103 & 3.199 & $-$3.780 & $-$3.900 & $-$3.790$^\ast$ & ---- & ---- & ---- & ---- & ---- & ---- & 19.0 & 6.66 & 25.7 & 7.33 \\
\hbox{Fe II} & 6149.246 & 3.890 & $-$2.720 & $-$2.800 & $-$2.690$^\ast$ & ---- & ---- & ---- & ---- & 11.2 & 6.18 & 22.9 & 6.48 & 18.4 & 6.74 \\
\hbox{Fe II} & 6247.559 & 3.892 & $-$2.310 & $-$2.400 & $-$2.300$^\ast$ & ---- & ---- & 45.3 & 7.20 & 31.9 & 6.51 & 37.6 & 6.51 & 38.3 & 6.99 \\
\hbox{Fe II} & 6416.930 & 3.892 & $-$2.650 & $-$2.900 & $-$2.640$^\ast$ & ---- & ---- & ---- & ---- & ---- & ---- & 27.7 & 6.58 & 28.3 & 7.04 \\
\hbox{Fe II} & 6432.677 & 2.891 & $-$3.520 & $-$3.500 & $-$3.570$^\ast$ & 42.2 & 6.71 & 35.3 & 7.10 & 33.0 & 6.69 & 33.6 & 6.51 & 38.5 & 7.12 \\
\hbox{Fe II} & 6456.380 & 3.903 & $-$2.100 & $-$2.200 & $-$2.050$^\ast$ & 26.5 & 5.95 & 32.7 & 6.62 & 38.5 & 6.45 & 43.7 & 6.43 & 50.4 & 7.08 \\
\hbox{Fe II} & 6516.077 & 2.891 & $-$3.320 & $-$3.370 & $-$3.310$^\ast$ & 43.2 & 6.47 & 29.0 & 6.67 & 35.2 & 6.49 & 51.0 & 6.67 & 50.9 & 7.17 \\
\noalign{\smallskip}
\hline
\end{longtable}
\tablebib{$^\ast$: Mel\'endez \& Barbuy (2009).}


\scalefont{1.45}
\begin{longtable}{cccrrrrrr}
\caption{\label{linelist} List of lines used in the present analysis, with the individual abundances.}\\
\hline\hline
\noalign{\smallskip}
\hbox{Species} &  \hbox{$\lambda$({\rm\AA})} & \hbox{$\chi$$_{ex}$(eV)}&
\hbox{log $gf$} & \hbox{A(X)} & \hbox{A(X)} & \hbox{A(X)} & \hbox{A(X)} & \hbox{A(X)}  \\
\hbox{} & \hbox{} & \hbox{} & \hbox{} & \hbox{221} & \hbox{224} & \hbox{230} & \hbox{235} & \hbox{238} \\
\hline
\noalign{\smallskip}
\endfirsthead
\caption{continued.}\\
\hline\hline
\noalign{\smallskip}
\hbox{Species} & \hbox{$\lambda$({\rm\AA})} &  \hbox{$\chi$$_{ex}$(eV)}  & \hbox{log$gf$} & \hbox{A(X)} & \hbox{A(X)} & \hbox{A(X)} & \hbox{A(X)} & \hbox{A(X)} \\
\hbox{} & \hbox{} & \hbox{} & \hbox{} & \hbox{221} & \hbox{224} & \hbox{230} & \hbox{235} & \hbox{238} \\
\hline
\noalign{\smallskip}
\endhead
\hline
\endfoot
\hbox{[O I]}   & 6300.311 & 0.000 & $-$9.716 & $+$8.40 & ------- & $+$8.25 & $+$8.40 & $+$8.85 \\
\hbox{Na I}  & 4982.813 & 2.104 & $-$0.962 & ------- & $+$5.45 & ------- & ------- & ------- \\
\hbox{Na I}  & 5688.205 & 2.104 & $-$0.450 & ------- & ------- & $+$4.70 & $+$5.22 & $+$5.88 \\
\hbox{Na I}  & 6154.230 & 2.102 & $-$1.560 & $+$5.65 & ------- & ------- & $+$5.18 & ------- \\
\hbox{Na I}  & 6160.753 & 2.104 & $-$1.260 & $+$5.70 & $+$5.70 & $+$5.10 & $+$5.22 & $+$5.98 \\
\hbox{Mg I}  & 5528.405 & 4.346 & $-$0.498 & ------- & ------- & ------- & $+$6.90 & ------- \\
\hbox{Mg I}  & 6318.720 & 5.108 & $-$2.100 & $+$7.22 & $+$7.50 & $+$7.18 & $+$7.18 & $+$7.35 \\
\hbox{Mg I}  & 6319.242 & 5.110 & $-$2.360 & ------- & $+$7.50 & ------- & $+$7.30 & ------- \\
\hbox{Mg I}  & 6765.450 & 5.750 & $-$1.940 & ------- & ------- & $+$7.20 & ------- & $+$7.55 \\
\hbox{Al I}  & 6696.015 & 3.143 & $-$1.569 & $+$6.00 & $+$6.30 & $+$5.64 & $+$5.85 & $+$6.25 \\
\hbox{Al I}  & 6698.673 & 3.143 & $-$1.870 & $+$6.00 & $+$6.28 & $+$5.90 & $+$5.72 & $+$6.21 \\
\hbox{Si I}  & 5665.555 & 4.920 & $-$2.040 & ------- & ------- & $+$6.90 & ------- & ------- \\
\hbox{Si I}  & 5690.425 & 4.930 & $-$1.870 & $+$6.70 & ------- & ------- & $+$6.85 & $+$7.18 \\
\hbox{Si I}  & 5948.545 & 5.082 & $-$1.300 & ------- & $+$7.14 & ------- & $+$6.75 & $+$7.25 \\
\hbox{Si I}  & 6142.494 & 5.619 & $-$1.500 & ------- & $+$7.32 & $+$6.80 & $+$6.77 & $+$7.40 \\
\hbox{Si I}  & 6145.020 & 5.616 & $-$1.450 & $+$7.00 & ------- & $+$6.95 & $+$6.77 & $+$7.15 \\
\hbox{Si I}  & 6155.142 & 5.619 & $-$0.850 & $+$6.96 & $+$7.35 & $+$6.90 & $+$6.76 & ------- \\
\hbox{Si I}  & 6237.328 & 5.614 & $-$1.010 & ------- & $+$7.25 & ------- & ------- & ------- \\
\hbox{Si I}  & 6243.823 & 5.616 & $-$1.300 & ------- & $+$7.20 & $+$7.00 & ------- & $+$7.15 \\
\hbox{Si I}  & 6414.987 & 5.870 & $-$1.130 & $+$6.80 & ------- & $+$7.30 & ------- & ------- \\
\hbox{Si I}  & 6721.844 & 5.860 & $-$1.170 & ------- & ------- & ------- & $+$6.80 & ------- \\
\hbox{Ca I}  & 5601.277 & 2.526 & $-$0.520 & ------- & ------- & ------- & $+$5.70 & $+$6.00 \\
\hbox{Ca I}  & 5867.562 & 2.933 & $-$1.550 & ------- & $+$6.20 & ------- & $+$5.72 & $+$6.00 \\
\hbox{Ca I}  & 6102.723 & 1.879 & $-$0.793 & $+$5.75 & $+$6.00 & $+$5.40 & $+$5.70 & $+$6.22 \\
\hbox{Ca I}  & 6122.217 & 1.886 & $-$0.200 & ------- & ------- & ------- & $+$5.85 & $+$5.85 \\
\hbox{Ca I}  & 6156.030 & 2.521 & $-$2.390 & $+$6.00 & ------- & ------- & $+$5.55 & $+$6.10 \\
\hbox{Ca I}  & 6161.295 & 2.510 & $-$1.020 & ------- & $+$5.90 & $+$5.35 & ------- & $+$6.00 \\
\hbox{Ca I}  & 6162.167 & 1.899 & $-$0.090 & $+$5.50 & $+$5.80 & $+$5.40 & $+$5.70 & $+$6.00 \\
\hbox{Ca I}  & 6166.440 & 2.521 & $-$0.900 & ------- & ------- & $+$5.40 & ------- & ------- \\
\hbox{Ca I}  & 6169.044 & 2.523 & $-$0.540 & $+$5.50 & $+$5.85 & $+$5.40 & ------- & $+$5.95 \\
\hbox{Ca I}  & 6169.564 & 2.526 & $-$0.270 & $+$5.50 & $+$5.80 & $+$5.42 & ------- & $+$5.80 \\
\hbox{Ca I}  & 6439.080 & 2.526 & $+$0.300 & $+$5.75 & $+$5.90 & $+$5.50 & $+$5.85 & $+$6.15 \\
\hbox{Ca I}  & 6455.605 & 2.523 & $-$1.350 & $+$5.45 & $+$5.90 & $+$5.70 & $+$5.80 & $+$6.25 \\
\hbox{Ca I}  & 6462.567 & 2.523 & $+$0.262 & ------- & ------- & ------- & ------- & $+$5.90 \\
\hbox{Ca I}  & 6464.679 & 2.523 & $-$2.100 & $+$5.60 & ------- & ------- & $+$5.50 & $+$6.00 \\
\hbox{Ca I}  & 6471.668 & 2.526 & $-$0.590 & ------- & $+$6.00 & $+$5.42 & $+$5.90 & $+$6.30 \\
\hbox{Ca I}  & 6493.788 & 2.521 & $+$0.000 & $+$5.70 & $+$6.10 & ------- & $+$5.83 & ------- \\
\hbox{Ca I}  & 6499.654 & 2.523 & $-$0.850 & $+$5.60 & $+$5.90 & ------- & $+$5.90 & $+$6.28 \\
\hbox{Ca I}  & 6572.779 & 0.000 & $-$4.320 & $+$5.80 & $+$6.35 & ------- & $+$5.80 & $+$6.38 \\
\hbox{Ca I}  & 6717.687 & 2.709 & $-$0.610 & $+$5.90 & $+$6.45 & $+$5.65 & $+$5.70 & $+$6.48 \\
\hbox{Ti I}  & 5689.459 & 2.230 & $-$0.400 & ------- & $+$4.65 & ------- & $+$4.32 & $+$4.55 \\
\hbox{Ti I}  & 5866.449 & 1.067 & $-$0.840 & ------- & $+$4.70 & $+$4.00 & $+$4.45 & ------- \\
\hbox{Ti I}  & 5922.108 & 1.046 & $-$1.460 & $+$4.38 & $+$4.94 & $+$4.50 & $+$4.45 & $+$5.00 \\
\hbox{Ti I}  & 5941.750 & 1.053 & $-$1.530 & ------- & $+$4.50 & $+$4.50 & $+$4.40 & ------- \\
\hbox{Ti I}  & 5965.825 & 1.879 & $-$0.420 & $+$4.32 & $+$4.45 & $+$4.10 & $+$4.32 & $+$4.85 \\
\hbox{Ti I}  & 5978.539 & 1.873 & $-$0.530 & $+$4.45 & $+$4.75 & ------- & $+$4.36 & ------- \\
\hbox{Ti I}  & 6126.224 & 1.070 & $-$1.430 & $+$4.35 & $+$4.45 & $+$4.40 & $+$4.45 & $+$4.82 \\
\hbox{Ti I}  & 6258.110 & 1.440 & $-$0.360 & $+$4.25 & $+$4.55 & $+$4.25 & $+$4.20 & $+$5.00 \\
\hbox{Ti I}  & 6261.106 & 1.430 & $-$0.480 & $+$4.20 & $+$5.00 & $+$4.20 & $+$4.40 & $+$5.00 \\
\hbox{Ti I}  & 6336.113 & 1.440 & $-$1.740 & $+$4.60 & $+$4.80 & ------- & $+$4.35 & $+$4.85 \\
\hbox{Ti I}  & 6554.238 & 1.440 & $-$1.220 & $+$4.42 & $+$4.80 & $+$4.30 & $+$4.38 & $+$4.85 \\
\hbox{Ti I}  & 6556.077 & 1.460 & $-$1.070 & $+$4.35 & ------- & ------- & $+$4.46 & ------- \\
\hbox{Ti I}  & 6599.113 & 0.900 & $-$2.090 & $+$4.50 & $+$4.96 & ------- & $+$4.50 & $+$4.90 \\
\hbox{Ti I}  & 6743.127 & 0.900 & $-$1.730 & ------- & ------- & ------- & $+$4.45 & $+$4.87 \\
\hbox{Ti II} & 5336.771 & 1.582 & $-$1.700 & ------- & ------- & ------- & $+$4.40 & $+$4.75 \\
\hbox{Ti II} & 5381.021 & 1.566 & $-$2.080 & ------- & ------- & ------- & $+$4.40 & ------- \\
\hbox{Ti II} & 5418.751 & 1.582 & $-$2.130 & ------- & ------- & ------- & $+$4.30 & ------- \\
\hbox{Ti II} & 6491.580 & 2.060 & $-$2.100 & $+$4.42 & $+$4.95 & $+$4.35 & $+$4.43 & $+$4.88 \\
\hbox{Ti II} & 6559.576 & 2.050 & $-$2.350 & $+$4.60 & $+$4.75 & $+$4.46 & ------- & $+$4.95 \\
\hbox{Ti II} & 6606.970 & 2.060 & $-$2.850 & ------- & $+$4.70 & $+$4.20 & ------- & $+$4.70 \\
\hbox{Y  I}  & 6435.004 & 0.066 & $-$0.820 & $+$1.60 & $+$1.95 & $+$1.90 & $+$1.60 & $+$1.92 \\
\hbox{Zr I}  & 6127.475 & 0.154 & $-$1.050 & $+$2.50 & ------- & $+$2.40 & $+$2.38 & ------- \\
\hbox{Zr I}  & 6134.585 & 0.000 & $-$1.280 & ------- & ------- & ------- & $+$2.40 & $+$2.55 \\
\hbox{Zr I}  & 6143.252 & 0.071 & $-$1.100 & ------- & $+$2.60 & ------- & $+$2.20 & $+$2.65 \\
\hbox{Ba II} & 6141.713 & 0.704 & $+$0.000 & $+$1.50 & $+$1.80 & $+$1.25 & $+$1.80 & $+$2.00 \\
\hbox{Ba II} & 6496.897 & 0.604 & $-$0.320 & $+$1.70 & ------- & $+$2.00 & $+$2.00 & $+$2.20 \\
\hbox{La II} & 6320.376 & 0.170 & $-$1.560 & $+$0.30 & $+$0.85 & ------- & $+$0.45 & $+$0.90 \\
\hbox{La II} & 6390.477 & 0.321 & $-$1.410 & ------- & $+$1.20 & ------- & $+$0.70 & $+$0.80 \\
\hbox{La II} & 6774.268 & 0.126 & $-$1.708 & ------- & ------- & ------- & $+$0.90 & ------- \\
\hbox{Eu II} & 6437.640 & 1.320 & $-$0.320 & $+$0.00 & $+$0.35 & $+$0.00 & $+$0.00 & $+$0.56 \\
\hbox{Eu II} & 6645.064 & 1.380 & $+$0.120 & ------- & ------- & $-$0.20 & $+$0.10 & ------- \\
\end{longtable}

\end{appendix}

\end{document}